\documentclass[12pt]{article}%
\pdfoutput=1

\usepackage{cite}
\usepackage{mathtools}
\usepackage{amsfonts}
\usepackage{amssymb}
\usepackage{authblk}
\usepackage{bbm}
\usepackage{bm}
\usepackage{caption}
\usepackage{epsfig}
\usepackage{epstopdf}
\usepackage{heck}
\usepackage[margin=1in]{geometry}
\usepackage{tabularx}
\usepackage{xcolor}
\usepackage{graphicx}

\usepackage{tikz}
\usetikzlibrary{decorations.pathreplacing}
\usetikzlibrary{calc}
\newcommand{\tikzmark}[1]{\tikz[remember picture,overlay]\node[yshift=0.75em](#1){};}

\usepackage{hyperref}
\usepackage[nameinlink,capitalize]{cleveref}

\DeclarePairedDelimiter\abs{\lvert}{\rvert}
\DeclarePairedDelimiter\bra{\langle}{\rvert}
\DeclarePairedDelimiter\ket{\lvert}{\rangle}
\newcommand*{\braket}[2]{\ensuremath\langle#1\vert#2\rangle}
\newcommand*{\braopket}[3]{\ensuremath\langle#1\vert#2\vert#3\rangle}

\makeatletter
\newcommand*{\diff}{\@ifnextchar^{\DIFF}{\DIFF^{}}}
\def\DIFF^#1{\mathop{\mathrm{\mathstrut d}}\nolimits^{#1}\gobblespace}
\newcommand*{\bigdiff}{\@ifnextchar^{\BIGDIFF}{\BIGDIFF^{}}}
\def\BIGDIFF^#1{\mathop{\mathrm{\mathstrut \mathcal{D}}}\nolimits^{#1}\gobblespace}
\def\gobblespace{\futurelet\diffarg\opspace}
\def\opspace{%
	\let\DiffSpace\!%
	\ifx\diffarg(%
		\let\DiffSpace\relax
	\else
		\ifx\diffarg[%
			\let\DiffSpace\relax
		\else
			\ifx\diffarg\{%
				\let\DiffSpace\relax
			\fi
		\fi
	\fi
	\DiffSpace
}
\makeatother
\DeclareMathOperator{\AdS}{AdS}
\DeclareMathOperator{\CFT}{CFT}
\DeclareMathOperator{\DThree}{D3}
\DeclareMathOperator{\DSeven}{D7}
\DeclareMathOperator{\dist}{dist}
\newcommand*{\id}{\bm{\mathrm{id}}}
\DeclareMathOperator{\KThree}{K3}
\DeclareMathOperator{\SL}{SL}
\DeclareMathOperator{\SU}{SU}
\DeclareMathOperator{\Tr}{Tr}
\DeclareMathOperator{\U}{U}

\newcommand*{\V}[1]{\overrightarrow{#1}}

\newcommand*{\bC}{\ensuremath\mathbb{C}}
\newcommand*{\bH}{\ensuremath\mathbb{H}}
\newcommand*{\bI}{\ensuremath\mathbb{I}}
\newcommand*{\bO}{\ensuremath\mathbb{O}}
\newcommand*{\bP}{\ensuremath\mathbb{P}}
\newcommand*{\bR}{\ensuremath\mathbb{R}}
\newcommand*{\bS}{\ensuremath\mathbb{S}}
\newcommand*{\bZ}{\ensuremath\mathbb{Z}}

\newcommand*{\cH}{\ensuremath\mathcal{H}}
\newcommand*{\cL}{\ensuremath\mathcal{L}}
\newcommand*{\cM}{\ensuremath\mathcal{M}}
\newcommand*{\cN}{\ensuremath\mathcal{N}}
\newcommand*{\cO}{\ensuremath\mathcal{O}}

\newcommand*{\cX}{\ensuremath\mathcal{X}}

\newcommand*{\numStacks}{\ensuremath K}
\newcommand*{\stackIndex}{\ensuremath k}
\newcommand*{\numNSFives}{\ensuremath m}

\begin{document}

\date{November 2021}

\title{Disorder Averaging and its UV (Dis)Contents}

\institution{PENN}{\centerline{$^{1}$Department of Physics and Astronomy, University of Pennsylvania, Philadelphia, PA 19104, USA}}
\institution{NYU}{\centerline{$^{2}$Center for Cosmology and Particle Physics, New York University University, New York, NY 10003, USA}}

\authors{Jonathan J. Heckman\worksat{\PENN}\footnote{e-mail: \texttt{jheckman@sas.upenn.edu}},
Andrew P. Turner\worksat{\PENN}\footnote{e-mail: \texttt{turnerap@sas.upenn.edu}}, and
Xingyang Yu\worksat{\NYU}\footnote{e-mail: \texttt{xy1038@nyu.edu}}
}

\abstract{We present a stringy realization of
quantum field theory ensembles in $D \le 4$ spacetime dimensions,
thus realizing a disorder averaging over coupling constants.
When each member of the ensemble is a conformal field theory with a standard semi-classical
holographic dual of the same radius, the resulting bulk can be interpreted as a single asymptotically Anti-de Sitter space geometry with a distribution of boundary components joined by wormhole configurations, as dictated by the Hartle--Hawking wave function.
This provides a UV completion of a recent proposal by Marolf and Maxfield that there is a high-dimensional Hilbert space
for baby universes, but one that is compatible with the proposed Swampland constraints of McNamara and Vafa.
This is possible because our construction is really an approximation that breaks down both at short distances,
but also at low energies for objects with a large number of microstates. The construction thus provides an explicit set of counterexamples
to various claims in the literature that holographic and effective field theory considerations can be
reliably developed without reference to any UV completion.}

\maketitle

\setcounter{tocdepth}{2}

\tableofcontents

\section{Introduction}

Disorder averaging is a well-defined procedure in any parametric family of quantum field theories (QFTs). Operationally, one
fixes the parameters of the theory, computes correlation functions for this choice, and then at the end, performs a
further average over a \emph{classical} probability distribution for these parameters. This is clearly a useful tool
for gaining access to ``typical'' behavior in various systems with a high degree of complexity (see, e.g., \cite{Anderson:1958vr,Sachdev:1992fk,Parisi:1992tk}).
It has also appeared in the context of holography (see, e.g., \cite{Hartnoll:2014cua, Aharony:2015aea, kitaev, Maldacena:2016hyu,Afkhami-Jeddi:2020ezh,Maloney:2020nni,Benjamin:2021wzr}) as well as other areas within high energy theory \cite{Rothstein:2012hk,Green:2014xqa,Craig:2017ppp, Balasubramanian:2020lux}.

Indeed, recent bottom-up considerations suggest that combining the principles of holography with effective field theory
in the context of the Euclidean gravitational path integral naturally results in the appearance of disorder averaging in the conformal field theory (CFT) on the boundary (see, e.g., \cite{Maldacena:2016hyu,Maldacena:2017axo,Penington:2019kki,Saad:2019lba,Stanford:2019vob,Marolf:2020xie} as well as \cite{Giddings:1989bq,Giddings:2020yes,Cottrell:2018ash,Sonner:2017hxc}).
One particularly surprising aspect of these considerations is that, at present, they have resisted an embedding in
string theory.\footnote{Other examples include the case of ``double holography'' \cite{Almheiri:2019hni}, which involves coupling a large $N$ gauge theory to gravity. For massless gravity, this is in sharp contradiction with all known string constructions and Swampland considerations \cite{Ooguri:2006in,Heckman:2019bzm}, but for a massive graviton, it might be possible \cite{Geng:2020qvw,Uhlemann:2021nhu,Geng:2021hlu,Raju:2021lwh}.} It is not hard to see that finding such a completion might be difficult, because the presence of an explicit classical distribution would seem to require treating gravity as an open system.

The hope, then, is that constraints imposed from requiring a UV completion can be sufficiently decoupled from long distance effects in the putative gravity dual. That being said, it is not entirely obvious that this is really the case. For example, in \cite{McNamara:2020uza} it was argued that the Swampland cobordism conjecture (see \cite{McNamara:2019rup}) precludes the existence of an AdS/CFT correspondence with ensemble averaging for $D > 2$ boundary systems, and in the case of $D \le 2$, the corresponding systems should be viewed as UV completed in a higher-dimensional system. A closely related point is that for many of these considerations, it is actually quite important that the couplings have no position dependence at all. This is required to have a consistent interpretation in terms of the creation of baby universes.\footnote{This can be seen as a consequence of Gauss's law for translation invariance; an emitted baby universe cannot carry energy or momentum because it has no boundary. We thank H.~Maxfield for helpful correspondence on this point.}

Our aim in this paper will be to engineer ensemble averaged QFTs embedded in string theory.
Let us state at the outset that aesthetically, the construction we present is a rather contrived UV completion.
That being said, it provides us with a general framework for testing the claim that disorder averaging in the context of
holography can truly be decoupled from stringy considerations.

The main idea behind our construction is to produce QFTs decoupled from gravity via open strings
localized on a small patch of the compactification geometry. To get a statistical ensemble, we simply consider multiple stacks of branes
at different locations in the transverse extra dimensions. In particular, by varying the profile of non-normalizable modes
in these directions, we can realize different low energy effective field theories with \emph{identical} field content but with
different values of the physical parameters. Given a set of $\numStacks$ stacks with couplings $\lambda_\stackIndex$ for $\stackIndex = 1, \dotsc, \numStacks$, we can consider a special class of operators $O_\stackIndex$ for $\stackIndex = 1, \dotsc, \numStacks$ given by ``tracing'' over all the different stacks:
\begin{equation}
	\bO \equiv O_1 + \dotsb + O_\numStacks\,.
\end{equation}
The key point is that the correlation functions of these $\bO$s factorize to leading order:
\begin{equation}\label{eq:Ofactor}
	\langle \bO(x) \bO'(y) \rangle \approx \sum_{1 \le \stackIndex \le \numStacks} \langle O_\stackIndex(x) O'_\stackIndex(y) \rangle\,,
\end{equation}
which in turn leads to an averaging over couplings. At short distances this approximation breaks down because we become sensitive to massive excitations that have been integrated out to reach the effective field theory in the first place.

Depending on the number density of stacks with a given value of $\lambda_\stackIndex$, it is clear that this is building up a ``binned'' version of an ensemble average. Provided we can engineer a suitable internal profile for the couplings and populate the stacks at distinct values of the couplings, it would appear that we can produce a large class of probability distributions $p(\lambda)$ for ensemble averaging. Note also that taking $\numStacks \gg 1$ provides a general way to get a good approximation by the binned distribution of its idealized smooth counterpart.

A particularly important special case is provided by requiring each
stack to realize a conventional holographic dual with the same value
of the bulk cosmological constant. In this case, we observe that the correlators
for the operators $\bO(x)$ reconstruct a \emph{single}
anti-de Sitter space (AdS) throat region. In this context, the appearance of a disorder average means that asymptotically,
we do not restrict ourselves to a fixed number of boundary components, but allow this to fluctuate, much as in \cite{Marolf:2020xie}.
Provided we only consider a number of boundary components much smaller than $\numStacks$, we thus make contact with the proposal of Marolf and Maxfield \cite{Marolf:2020xie}, and in the limit where $\numStacks$ is very large, this provides an adequate way to build up an ensemble average and its holographic dual.

However, with an actual construction in hand, we can also identify two general ways in which our UV completion breaks down.
First of all, we clearly have a large number of sequestered stacks of branes,
so if we proceed to higher energies, we should observe additional contributions beyond the
approximate factorization appearing in \cref{eq:Ofactor}. This is not altogether surprising, but already points to the fact
that the UV completion does place an intrinsic limitation on the sorts of correlation functions we can consider.

Perhaps more surprisingly, there is \emph{another} way in which the approximation can break down, and it is something that
occurs even if we restrict to observables deep in the infrared. This is due to the fact that our UV completion implicitly makes reference to a fixed $\numStacks$, and we can actually distinguish between our binned approximation and a smooth distribution after sampling $n_*$ times (see \cref{eq:nstar} later on for the precise definition). In particular, if we consider any bulk object characterized
by a density matrix with order $n_*$ or more entries (as, for example, we would need to
discuss in constructing the Page curve of a macroscopic black hole \cite{Page:1993wv}, see, e.g., \cite{Page:2013dx}),
then our putative ensemble average has been pushed beyond its regime of validity.

To make these considerations precise, we present a number of examples illustrating
how to generate ensemble averaging for appropriate string-based constructions. Curiously enough, the case where we can maintain
the most control is for $D = 4$ superconformal field theories (SCFTs) with an ensemble average over marginal coupling constants.
We illustrate this both in terms of compactifications of 6D little string theories (LSTs), brane box configurations, and
D3-brane probes of orbifold singularities. In this case, the extra dimensions transverse to the brane stacks
provide enough flexibility to produce a nearly arbitrary probability distribution with support on
the moduli space of marginal couplings. As another class of examples, we consider various $D = 2$ SCFTs obtained in a similar fashion.
In this case, we find that for examples where we can reliably extract an $\AdS_3$ dual description, it is often simplest to consider the fibration of a Calabi--Yau $n$-fold over a subvariety of its moduli space, resulting in a higher-dimensional (non-compact) Calabi--Yau geometry (see \cref{sec:FIBRATION}). The limitation of this sort of construction is then that our ensemble averages are necessarily restricted to a particular subset of moduli. Similar considerations hold for $D = 1$ ``SCFTs'' of the sort that appear in the construction of 4D black holes obtained from type II strings on Calabi--Yau threefold backgrounds. Here, we again get an ensemble average, as associated with $\AdS_2$ vacua. It is worth pointing out that at no point do we truly get a 2D gravitational theory; rather we get an $\AdS_2 \times S^2 \times X_6$ background.

Our method meets with less success in the case of $\AdS_4$, $\AdS_6$, and $\AdS_7$ vacua, as associated with 3D, 5D, and 6D SCFTs. In the case of 3D SCFTs, this may just be a failure of imagination / stamina. In the case of $D > 4$ SCFTs, we face the fact that there are no marginal deformations available that preserve supersymmetry \cite{Louis:2015mka, Cordova:2016xhm}.

The rest of this paper is organized as follows. In \cref{sec:GENERAL}, we present some more details on the general idea of our construction, and its regime of validity. We present a holographic interpretation of this construction in \cref{sec:HOLO}. \Cref{sec:EXAMPLES} presents explicit string-based examples. We present a brief discussion in \cref{sec:DISC}.
Some additional technical details are deferred to \cref{sec:FIBRATION}.

\section{Engineering an Ensemble}\label{sec:GENERAL}

In this \namecref{sec:GENERAL}, we present the main idea of generating a QFT ensemble average in string theory constructions,\footnote{For the reader interested in learning more about string theory, the authors recommend~\cite{Zwiebach:2004tj, Ibanez:2012zz}. It is a fascinating subject.} giving explicit examples later in \cref{sec:EXAMPLES}. We are specifically interested in the case where the coupling constants of the QFT are truly constant in the sense that they have no spacetime dependence. This is the case that has been of primary interest in recent holographic investigations, and also turns out to be the most challenging one to engineer in the context of string constructions. See reference \cite{Balasubramanian:2020lux} for how to get an ensemble average in a gravitational system but with spacetime-dependent couplings.

Our goal will be to mimic the effects of disorder averaging in a QFT via a string construction. More precisely, we suppose that our QFT depends on a set of couplings $\{\lambda\} \equiv \lambda$, and that we have a smooth classical distribution $p_\text{smooth}(\lambda)$. Given operators $O^{(1)}, \dotsc, O^{(m)}$ of the QFT, the disorder averaged correlator is obtained by evaluating the correlation function with respect to a fixed value of $\lambda$, and then performing a further averaging with respect to $p_\text{smooth}(\lambda)$:
\begin{equation}
	\overline{\langle O^{(1)} \dotsm O^{(m)} \rangle} \equiv \int \diff{\lambda} \; p_\text{smooth}(\lambda) \langle O^{(1)} \dotsm O^{(m)} \rangle\,.
\end{equation}

Our plan will be to engineer an ensemble of QFTs that are decoupled from gravity. Loosely speaking, this is arranged by taking a limit in which we try to retain some localized degrees of freedom while decoupling the gravitational degrees of freedom.\footnote{For example, open strings attached to a stack of D-branes.} This in turn means that some degrees of freedom are non-normalizable, and these descend to coupling constants $\{\lambda\}$ of the QFT sector. An important comment here is that even though these degrees of freedom are constants in $D < 10$ spacetime dimensions, they are still dynamical in the full $D = 10$ (resp., $D = 11$) spacetime associated with string theory (resp., M-theory).\footnote{Indeed, in reference \cite{Balasubramanian:2020lux} it was noted that in a string compactification with multiple QFT sectors coupled to gravity, performing a partial trace over all but one such QFT sector results in a system characterized by a mixed state for position-dependent coupling constants. A subtlety with this approach is that the process of decoupling gravity also tends to force the previously obtained mixed state back into a pure state, and we will ultimately need to take a different tack to generate a position-independent ensemble average.}

\begin{figure}[t!]
	\centering

	\includegraphics[width=15.5cm]{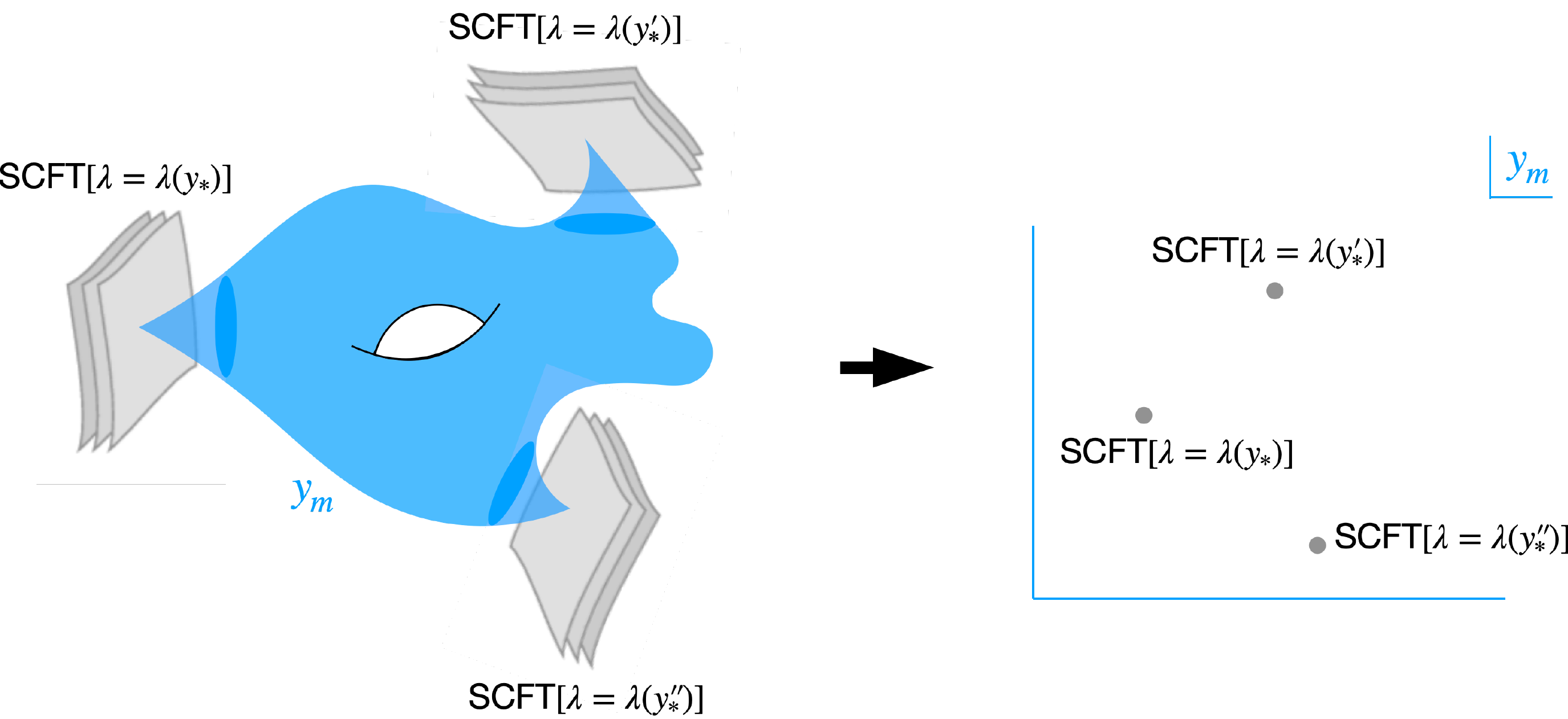}

	\caption{Copies of the same QFT sector separated from one another in the extra dimensions. QFTs are engineered on the worldvolume of branes (colored grey). The extra-dimensional geometry is colored blue.}
	\label{fig:QFT sample in string theory}
\end{figure}

Now, precisely because we are dealing with a non-compact extra-dimensional geometry, we are free to consider multiple copies of the same QFT sector, just separated from one another in the extra dimensions. This stringy setup is roughly illustrated in \cref{fig:QFT sample in string theory}.
 The parameters on each local sector are then specified by the internal profile of the non-normalizable modes in the vicinity of each local model. Introducing $\numStacks$ such local sectors, we can label the corresponding couplings as $\lambda_{\stackIndex}$ for $\stackIndex = 1, \dotsc, \numStacks$, where $\lambda_\stackIndex \equiv \{\lambda_\stackIndex\}$ denotes the full set of couplings in each local model. Since the field content is assumed to be identical, we also have the same set of operators $\{O_\stackIndex\}$ for each local model. The different sectors are decoupled only in an approximate sense because we can also consider degrees of freedom that stretch from one sector to the other (e.g., open strings in D-brane models), and this introduces a UV cutoff $\Lambda_\text{UV}$ for this approximate factorization of the Hilbert space of states:
\begin{equation}
	\bH \approx \cH_1 \otimes \dotsb \otimes \cH_\numStacks\,.
\end{equation}
In the effective action for the full system, the action breaks up into a set of distinct contributions that then mix via higher-dimension operators with suppression scale $\Lambda_\text{UV}$:
\begin{equation}
	\bS = S_1 + \dotsb + S_\numStacks + S_\text{mix}\,.
\end{equation}
Given a specific operator of a single sector, there is a natural subset of operators obtained
by forming a sum over all such sectors:
\begin{equation}\label{eq:otown}
	\bO \equiv O_1 + \dotsb + O_\numStacks\,.
\end{equation}
Normalized correlation functions for $n$ such operators are then evaluated as:
\begin{equation}
	\langle \bO^{(1)} \dotsm \bO^{(n)} \rangle_\text{normalized} \equiv \frac{1}{\numStacks} \langle \bO^{(1)} \dotsm \bO^{(n)} \rangle_{\bH}\,,
\end{equation}
where on the righthand side the correlation function is evaluated with respect to the ground state of the full system. The pre-factor of $1 / \numStacks$ can be understood as the statement that we just require a well-behaved large $\numStacks$ limit. Another way to understand the same requirement is that we are just measuring all correlation functions in units of the traced identity operator:
\begin{equation}
	\bI = \id_1 + \dotsb + \id_\numStacks\,.
\end{equation}

Correlation functions for the $\bO$ operators close to leading order. To see why, we note that:
\begin{equation}
	\langle \bO^{(1)} \dotsm \bO^{(n)} \rangle_\text{normalized} \approx \frac{1}{\numStacks} \sum_{1 \le \stackIndex \le \numStacks} \langle O_\stackIndex^{(1)} \dotsm O_\stackIndex^{(n)} \rangle_{\cH_\stackIndex}\,.
\end{equation}
where to leading order the different factors $\cH_i$ are decoupled from one another. There can also be cross-terms between
the stacks, but these are all subleading contributions. Consider first connected correlators. In this case, sequestering the
stacks suppresses such contributions. For disconnected correlators (for example $\langle O_k O_k \rangle \langle O_l O_l \rangle$ with $k \ne l$), we further note that such contributions are kinematically suppressed relative to their connected counterparts
(as is clear by passing to momentum space).\footnote{We thank H.~Maxfield for correspondence on this point.}
In this sense it is a consistent truncation.

Our main claim is that this can be used to build up a discretized approximation to a disorder averaging by a continuous distribution $p_\text{smooth}(\lambda)$. To see why, we observe that each expectation value over $\cH_\stackIndex$ makes reference to the couplings
$\{\lambda_\stackIndex\}$ of that sector. If we have a total of $\numStacks(\lambda)$ sectors with a particular set of couplings, then the probability associated with this choice is:
\begin{equation}\label{eq:discDist}
	p_\text{disc}(\lambda) = \frac{\numStacks(\lambda)}{\numStacks}\,.
\end{equation}
As it stands, this is a discrete probability distribution, but it is important to note that in any actual string construction, there is an intrinsic ``spread'', as associated with the overall tension of a brane/region of localization in the internal geometry.\footnote{For example, there is a minimal length scale $\ell_\text{min} \sim \left(\frac{1}{T_p}\right)^{\frac{1}{p + 1}}$ that a D$p$-brane of tension $T_p \sim (g_s\ell_\text{st}^{p + 1})^{-1}$ can probe (see, e.g., \cite{Shenker:1995xq}).} For this reason, it is more appropriate to view our construction as building up a continuous distribution, but one that has been suitably ``binned''. More precisely, introducing an indicator function $I_{\lambda', \varepsilon_{\lambda'}}(\lambda)$, which has unit area and has support on a small region of size $\varepsilon_{\lambda'}$ centered around $\lambda'$, the probability of drawing $\lambda$ builds up a histogram comprised of small bins of size $\varepsilon_{\lambda'}$:
\begin{equation}\label{eq:binnedDist}
	p_\text{bin}(\lambda) = \sum_{\lambda'} I_{\lambda' , \varepsilon_{\lambda'}}({\lambda}) \frac{\numStacks(\lambda')}{\numStacks}\,.
\end{equation}
The values of the $\varepsilon_{\lambda'}$ depend on the UV cutoff $\Lambda_\text{UV}$, as well as specific details of the model and construction.

The approximation just developed enables us to closely mimic the disorder average we would get for a smooth probability distribution, but there are clearly some limitations. One of these is already apparent from the general setup: in the correlation functions for the $\bO$ operators, we observe that factorization will begin to break down when any correlators probe a short distance scale of size $\Lambda_\text{UV}$. Indeed, this is just the statement that we have a UV cutoff. Provided we work at long distances close to an infrared fixed point, we can hope to neglect such contributions.

There is also an entropic breakdown as associated with sampling the distribution $p_\text{bin}(\lambda)$ a large number of times. To see the issue, we consider two sorts of observers, a ``daemonic observer'' who has access to the full set of operators $O_i$, and another ``ignorant observer'' who is confined to making do with just the $\bO$ operators.\footnote{This is related to the broader question of how well a low energy observer can ever hope to reconstruct a given UV completion, see, e.g., \cite{Heckman:2013kza,
Balasubramanian:2014bfa, Heckman:2016wte, Heckman:2016jud, Fowler:2020rkl, Balasubramanian:2020lux}
as well as \cite{Balasubramanian:1996bn, mehta2014exact, Erdmenger:2020vmo, Halverson:2020trp,
Stout:2021ubb, Erdmenger:2021sot, Erbin:2021kqf, Fowler:2021oje}.} For the daemonic observer, they can, after performing $n$ measurements associated with correlations between the $O_i$ and the $\bO$ operators, extract $n$ independent and identically distributed (iid) draws from $p_\text{bin}$. The discrepancy between $p_\text{smooth}(\lambda)$ and $p_\text{bin}(\lambda)$ is captured by the relative entropy/Kullback--Leibler divergence, which reflects the amount of information we would gain upon switching from the binned distribution $p_\text{bin}(\lambda)$ to the ``true'' (although, from our perspective, unobtainable) distribution $p_\text{smooth}$:
\begin{equation}\label{eq:nstar}
	D_\text{KL}(p_\text{smooth} \mid\mid p_\text{bin}) = \int \diff{\lambda} \; p_\text{smooth}(\lambda) \log\frac{p_\text{smooth}(\lambda)}{p_\text{bin}(\lambda)} \approx c \varepsilon^2 \equiv \frac{1}{n_*}\,,
\end{equation}
where $c$ an order one constant dependent on the details of the distribution, and $\varepsilon$ is a representative value of the
size of our histogram bins. We can view $\varepsilon^{-1}$ as telling us the total number of distinct histogram bins. An observer who samples the distribution order $n \sim n_*$ times will be able to detect the difference between the string construction and the smooth ``idealization''. Let us note that typically $\varepsilon^{-1} \ll \numStacks$, since we necessarily need to group multiple values of the couplings in a single histogram bin to get an adequate approximation of $p_\text{smooth}(\lambda)$ in the first place.

Similar considerations hold for the ``ignorant observer,'' but their strategy for inferring the existence of a distribution over couplings is somewhat different. In this case, the point is to sample a number of different correlation functions over just the $\bO$s, and in so doing infer the higher moments of the probability distribution $p_{\text{bin}}(\lambda)$. Here, distinguishability is governed by the number of moments of the distribution they are able to extract.

While the details of a particular model will dictate the specific operators to consider,
we can always consider the partition function in Euclidean signature, as obtained by placing
our QFT on some $D$-dimensional background. For a fixed value of the couplings $\lambda$, we have,
in each of our local sector path integrals:
\begin{equation}
	Z_\lambda = \int \bigdiff{\phi} \; e^{-S_\lambda[\phi]}\,,
\end{equation}
in the standard notation. Observe that in our stringy construction,
the quantity $\overline{Z}$ is obtained from just performing a sum over each individual local sector:
\begin{equation}\label{eq:Zover}
	\overline{Z} = \frac{1}{\numStacks} \sum_\numStacks \int \bigdiff{\phi_\stackIndex} \; e^{-S_{\lambda_\stackIndex}[\phi_\stackIndex]}\,.
\end{equation}
Once we engineer $p_\text{bin}(\lambda)$, we also implicitly get a probability distribution over just $Z$. To see why, we can similarly consider the quantities $Z_{\stackIndex}, Z_{\stackIndex}^2, \dotsc$, and compute the corresponding moments for the partition function.

In terms of our $\bO$ operator formalism, we construct the corresponding operators as follows.
First of all, we build a copied unnormalized thermal density matrix:
\begin{equation}
\bP^{(1)} \equiv \rho_1 + \dotsb + \rho_\numStacks\,,
\end{equation}
with $\rho_\stackIndex \equiv \exp(-\beta H_\stackIndex) \otimes \rho_{\text{GND}, \stackIndex_\perp}$. Here,
$H_\stackIndex$ is the Hamiltonian on the $\stackIndex^\text{th}$ stack and we explicitly tensor by $\rho_{\text{GND}, \stackIndex_\perp}$,
the ground state associated with all the other Hilbert space factors, since as far as the evaluation of the partition function is
concerned, an observer there has no access to any other states.\footnote{It is of course tempting to
also consider a different class of operators defined as $\cO = \sum_\stackIndex O_\stackIndex \otimes \rho_{\text{GND}, \stackIndex_\perp}$,
which share many of the same properties as the $\bO$ operators. One issue is that it is physically rather awkward to
create an excitation in one stack, and simultaneously enforce a projection onto the ground state of the other stacks.
As already mentioned, however, this is quite appropriate in constructing a corresponding partition function.}
Upon evaluating $\numStacks^{-1} \Tr\bP^{(1)}$, we then get
just $\overline{Z}$, as in \cref{eq:Zover}. To get the higher order terms such as $Z_\stackIndex^m$,
we need a corresponding operator acting on a Hilbert space of states, so we
consider $m$ replicas of $\cH_\stackIndex$, namely the $m$-fold tensor product $\cH_\stackIndex^{\otimes m}$.
With respect to this, we introduce the replica density matrix
for the $\stackIndex^\text{th}$ stack:
\begin{equation}
\rho^{(m, \text{rep})}_\stackIndex \equiv \rho^{(1)}_\stackIndex \rho^{(2)}_\stackIndex \dotsm \rho^{(m - 1)}_\stackIndex \rho^{(m)}_\stackIndex\,,
\end{equation}
and then the corresponding copied thermal density matrix including all the replicas is:
\begin{equation}
\bP^{(m)} \equiv \rho^{(m, \text{rep})}_1 + \dotsb + \rho^{(m, \text{rep})}_\numStacks\,.
\end{equation}
Upon evaluating $\numStacks^{-1} \Tr \bP^{(m)}$, we then get
just $\overline{Z^m}$. Continuing in this fashion, we can clearly
treat the partition function itself as a random variable,
and evaluate its moments.

\section{Holographic Interpretation}\label{sec:HOLO}

Let us now specialize to the case where each local model is a conformal field theory with a semi-classical AdS gravity dual with the same bulk value of the cosmological constant. In this case, we clearly obtain a large number of AdS throats in which bulk fields $\Phi^\text{bulk}_\stackIndex $ have a boundary condition set by the particular values of the couplings on the boundary:
\begin{equation}
	\Phi^\text{bulk}_\stackIndex \to \lambda_\stackIndex\,.
\end{equation}
Our aim will be to understand the sense in which the construction just presented can be interpreted in terms of a \emph{single} AdS throat region. In the process, we will make contact with the baby universe interpretation of ensemble averaging proposed in \cite{Marolf:2020xie}, but one that respects the Swampland constraints of \cite{McNamara:2019rup}.

Our approach to this question will be to focus on the operator subsector defined by the $\bO$s of our QFT (now a CFT) sector. Along these lines, consider a local operator $\bO(x)$:
\begin{equation}
	\bO(x) = O_1(x) + \dotsb + O_\numStacks(x)\,,
\end{equation}
where each summand has the same field content on its respective stack.
By the standard rules of \cite{Gubser:1998bc}, we know that for each $O_\stackIndex(x)$, we can (in principle) write down a corresponding bulk field profile in an $\AdS_{(\stackIndex)}$ geometry. Said differently, each bulk quantity ``casts a shadow'' corresponding to a specific dual in the CFT. On the other hand, precisely because the $O_\stackIndex$ are built from the same fields, we see that even though we are dealing with a large number of AdS throats, the collective motion described by $\bO(x)$ only accesses a single AdS. Said differently, because the connected correlators for the $\bO$ operators close, according to the standard AdS/CFT dictionary \cite{Gubser:1998bc}, they reconstruct a single AdS throat.

Another way to arrive at the same conclusion is to consider the geometric entanglement entropy for the ground state between a ball $B$ and its complement $B^c$. Again, if we were initially dealing with a single local sector of our construction, we would simply introduce the pure state $\rho_\stackIndex = \ket{0}_\stackIndex \, _\stackIndex\bra{0}$. As is well known, this has a gravity dual description in terms of a ``minimal area surface'' homologous to $B$, and its ``area'' tracks with the entanglement entropy \cite{Ryu:2006bv,Hubeny:2007xt}. In the present setting, the ground state is given by the tensor product:
\begin{equation}
	\rho_\text{GND} = \rho_1 \otimes \dotsb \otimes \rho_\numStacks\,.
\end{equation}
In the limit where the additional throats are sequestered from each other, the partial trace collapses to a single ``diagonal'' contribution. Again, the interpretation is that for this set of states, we are building up a single bulk ``minimal area surface''.

\begin{figure}
	\centering
	\includegraphics[width=12cm]{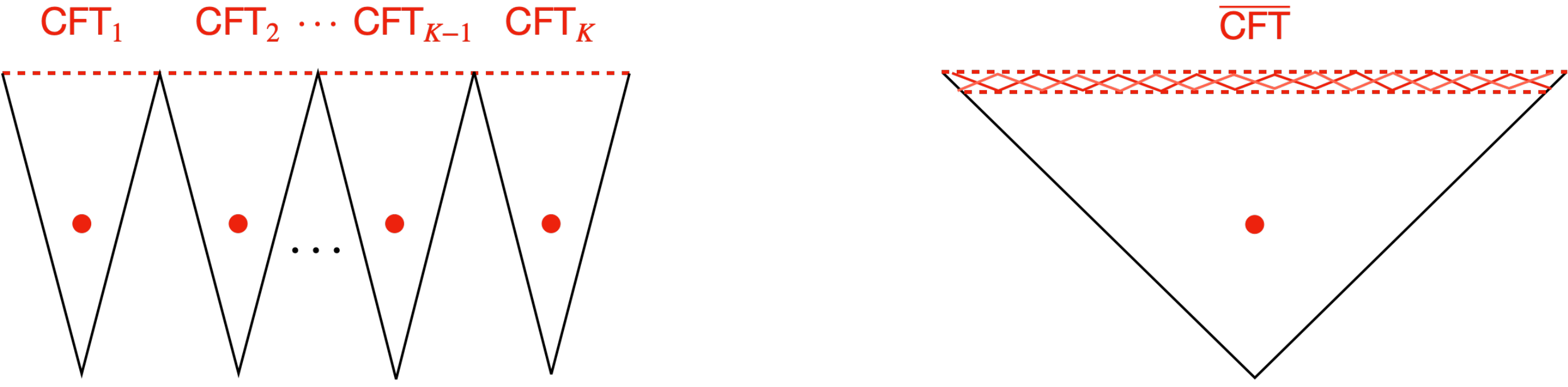}
	\caption{On the left, the string theory construction builds a large number of $D$-dimensional CFTs, each of which has its own dual AdS throat. The closed subset of operators defined by the $\bO$s, however, only reconstructs a single AdS throat, which is dual to an ensemble-averaged CFT, shown on the right.}
	\label{multi to single throat}
\end{figure}

Summarizing the discussion so far, we have seen that even though our stringy geometry is building up a large number of AdS throats,
the closed subsector defined by the $\bO$ operators only reconstructs a single throat, and this is the one that produces an ensemble-averaged CFT. In other words, only a single bulk AdS geometry is needed to match to the subsector associated with the $\bO$s. See Figure \ref{multi to single throat} for an illustrative depiction.

In the UV complete realization in terms of multiple AdS throats, one might of course ask whether there could be wormhole configurations that join these individual throats, perhaps via some generalization of the construction presented in \cite{Balasubramanian:2020ffd}. The general point is that our construction only mimics an ensemble average provided we have sufficient statistics. For this reason, it is natural to expect that the dominant contribution from saddle points of the Euclidean path integral instead comes from wormhole configurations that have fractionated, i.e., they join many boundaries.

Of course, the notion of ensemble averaging in AdS/CFT has recently been a topic of much interest, and so it is natural to ask how the present description fits with this. To this end, we next turn to a brief summary of the proposal of Marolf and Maxfield (MM) \cite{Marolf:2020xie} in terms of baby universes, and then explain why it can be a valid approximation compatible with the considerations of McNamara and Vafa (McV) \cite{McNamara:2019rup}.

\subsection{Baby Universe Disintegration}

To frame the discussion to follow, we first provide a brief summary of the MM proposal for how to capture the effects of ensemble averaging in AdS/CFT from the perspective of the gravity dual. Following \cite{Marolf:2020xie}, consider an AdS gravity theory with a set of fields denoted as $\Phi$ (including the metric), with boundary conditions labelled by $J$: $\Phi \sim J$. Note that the boundary can have more than one component generically. The Euclidean gravitational path integral defined by an asymptotic boundary with $n$ connected components is then
\begin{equation}\label{eq:def of gravity path integral}
	\langle Z[J_1] \dotsm Z[J_n] \rangle \equiv \int_{\Phi \sim J} \bigdiff{\Phi} \; e^{-S_\text{grav}[\Phi]}\,,
\end{equation}
where $J_1, \dotsc, J_n$ correspond to different components of the asymptotic boundary. As a point of notation, let us emphasize that here and throughout this subsection, $J_a$ really are just the boundary couplings, but to emphasize that they are \emph{not} associated with a particular set of stacks in our UV completion, we shall write $J_a$, with $a$ having no relation to the indexing of all the $\lambda_\stackIndex$.

\begin{figure}[t!]
	\centering

	\includegraphics[width=9cm]{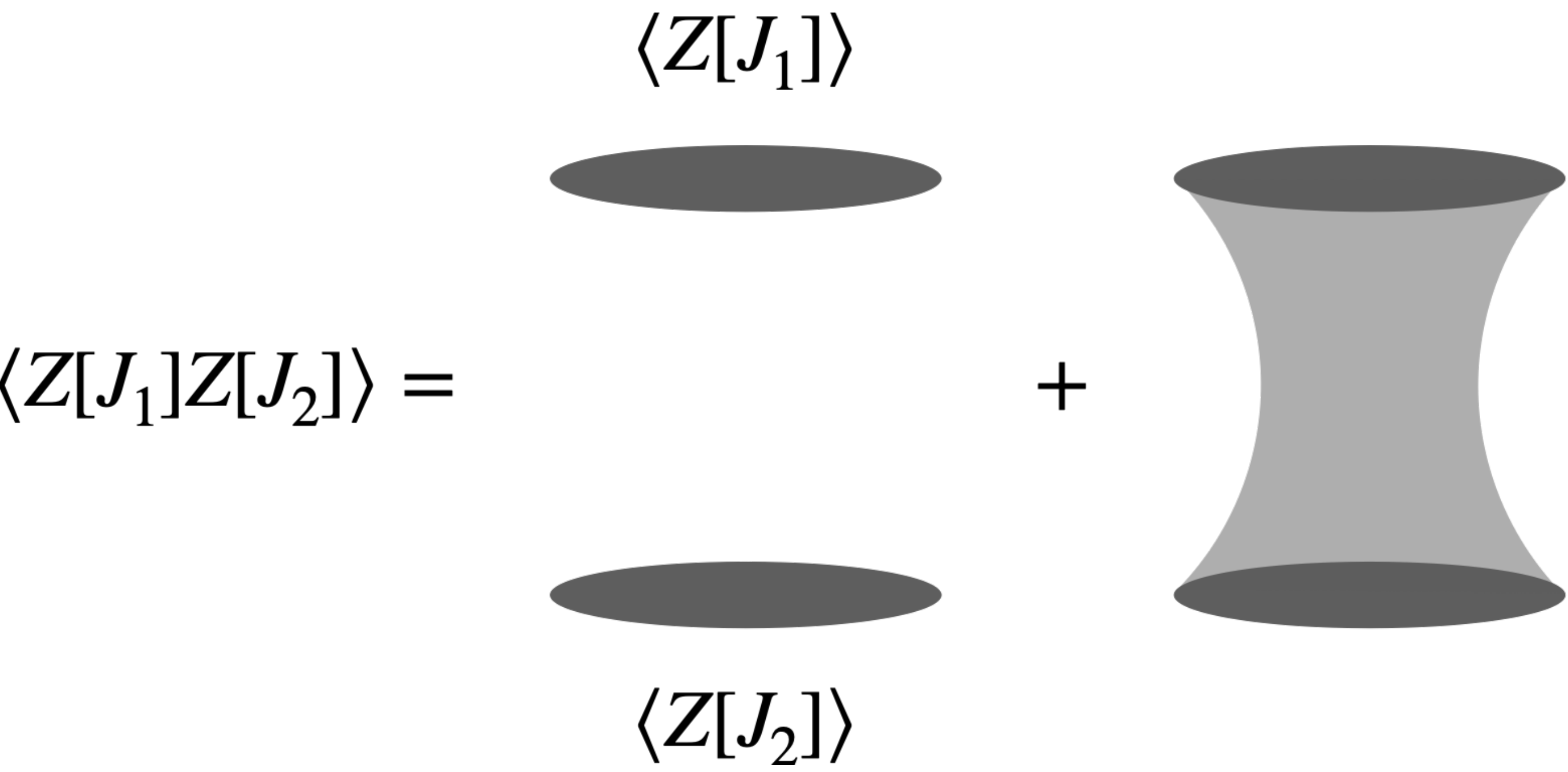}

	\caption{The presence of Euclidean wormholes results in non-factorization of the gravitational path integral.}
	\label{fig:factorization}
\end{figure}

The path integral defined in \cref{eq:def of gravity path integral} cannot generically be factorized into those of disconnected boundaries \cite{Witten:1999xp,Maldacena:2004rf}, e.g., with $n=2$
\begin{equation}
\langle Z[J_1] Z[J_2] \rangle \ne \langle Z[J_1] \rangle \langle Z[J_2] \rangle\,.
\end{equation}
This is because of the presence of Euclidean wormholes corresponding to the bulk manifold whose connected component includes the two boundaries. See \cref{fig:factorization}. From the bulk perspective, this non-factorization comes from the dynamical interactions of two boundaries connected by Euclidean wormholes. From the boundary-CFT point of view, nevertheless, this non-factorization should be rather interpreted as the ensemble average over a classical probability distribution as follows.

By cutting open the above integral carefully so that the intermediate slice intersects no asymptotically AdS boundaries, one defines the baby universe Hilbert space $\cH_\text{BU}$ for the complete set of intermediate states separating ``past'' and ``future''. The set of boundary conditions $\{J_1, \dotsc, J_n\}$ is then associated to a state
\begin{equation}
	\ket{Z[J_1] \dotsm Z[J_n]} \in \cH_\text{BU}\,.
\end{equation}
One special state is the Hartle--Hawking state with no boundary. Its norm gives rise to the ``cosmological partition function''
\begin{equation}
	\braket{\text{HH}}{\text{HH}} =\int_\text{no boundary} \bigdiff{\Phi} \; e^{-S_\text{grav}[\Phi]}\,.
\end{equation}

Next, introduce operators $\widehat{Z[J]}$ on $\cH_\text{BU}$ for any boundary condition $J$ so that
\begin{equation}
	\widehat{Z[J]} \ket{Z[J_1] \dotsm Z[J_n]} = \ket{Z[J] Z[J_1] \dotsm Z[J_n]}\,.
\end{equation}
The eigenstates of $\widehat{Z[J]}$ then form a basis of $\cH_\text{BU}$, and are defined by
\begin{equation}
	\widehat{Z[J]} \ket{\alpha} = Z_\alpha[J] \ket{\alpha} \quad \forall J\,,
\end{equation}
with $\braket{\alpha'}{\alpha} = \delta_{\alpha' \alpha}$.
Note that any boundary condition can be derived from the corresponding operators acting on Hartle--Hawking state as
\begin{equation}
	\ket{Z[J_1] \dotsm Z[J_n]} = \widehat{Z[J_1]} \dotsm \widehat{Z[J_n]} \ket{\text{HH}}\,.
\end{equation}
The gravitational path integral in \cref{eq:def of gravity path integral} can then be expressed as
\begin{equation}
\begin{aligned}
	\langle Z[J_1] \dotsm Z[J_n] \rangle &= \braopket{\text{HH}}{\widehat{Z[J_1]} \dotsm \widehat{Z[J_n]}}{\text{HH}} \\
	&= \sum_{\alpha_0, \dotsc, \alpha_n} \braket{\text{HH}}{\alpha_0} \braopket{\alpha_0}{Z[J_1]}{\alpha_1} \dotsm \braopket{\alpha_{n - 1}}{Z[J_n]}{\alpha_n} \braket{\alpha_{n}}{\text{HH}} \\
	&=\braket{\text{HH}}{\text{HH}} \sum_\alpha p_\alpha Z_\alpha[J_1] \dotsm Z_\alpha[J_n]\,,
\end{aligned}
\end{equation}
where $p_\alpha$ is the probability for each $\alpha$ state, computed by $p_\alpha = \frac{\abs{\braket{\text{HH}}{\alpha}}^2}{\braket{\text{HH}}{\text{HH}}}$.

The appearance of the $\ket{\alpha}$ states is rather disturbing, especially in the context of string theory where we have no evidence at all for such tunable parameters. Indeed, in string constructions coupled to gravity, all known examples of coupling constants descend from dynamical moduli. This point was significantly sharpened in \cite{McNamara:2019rup} where they showed that in even more general terms, the cobordism hypothesis of the Swampland program is enough to require the baby universe Hilbert space $\cH_\text{BU}$ to be one-dimensional, namely we can indeed speak of fixing the boundary values of the coupling constants, just as we would in ``standard'' AdS/CFT. From this perspective, ensemble averaging really has no general meaning in a UV complete framework such as string theory.

We now argue that in spite of appearances, our embedding in string theory provides a way to make contact with both proposals.

First of all, the very fact that we have an ensemble average means that any putative gravity dual will likely have to match on to the
characterization provided by the MM picture. Indeed, assuming we can work out the ``standard'' AdS dual for a single throat in our construction, we already know the bulk field content, and thus in principle can discuss the computation of the Hartle--Hawking wave function for this theory.

Note that the probability distribution $p_\text{bin}(\lambda)$ we engineer
in string theory does not directly correspond to the probability $p_\alpha$
for $\alpha$ states, though the two notions are clearly implicitly related in some way.
From the perspective of the bulk gravitational system, it is tempting
to say that the specific details of the path integral dictate a particular ``preferred choice'' for $p_\text{smooth}(\lambda)$
(see, e.g., \cite{Saad:2019lba, Stanford:2019vob, Marolf:2020xie}), although even this relies on having enough data in the form of a specific set of $\alpha$ states, and a specific choice of bulk gravitational action with which to construct the Hartle--Hawking state in the first place. Indeed, turning the discussion around, there seems to be little constraint on what sort of $p_\text{bin}(\lambda)$s we can end up generating,
and so we leave it as an interesting question to determine precisely how to fill in this entry of the AdS/CFT correspondence.
From the perspective of the present construction, we take this to mean that there are ambiguities in specifying the Euclidean path integral
for quantum gravity, and resolving these ambiguities in different ways can result in different choices for the ensemble average in the boundary theory.

Once we accept the existence of the $\ket{Z[J_1] \dotsm Z[J_n]}$ states, the appearance of the baby universe states $\ket{\alpha}$ would appear to follow. At this point, however, we recall that our ensemble average picture can break down, both at high energies, but also entropically whenever $n$ gets sufficiently large. Probing either regime of validity shatters the illusion and we can no longer work in terms of a single AdS throat region. For example, when the number of boundary components $n$ becomes sufficiently large, (i.e., of order $n_*$ of \cref{eq:discDist}), then we have already seen there can be an entropic breakdown. Observe that when $n$ is really of order $\numStacks$, we can even resolve the individual AdS throats, and so in this limit, we just recover the standard AdS/CFT dictionary, with a single one-dimensional baby universe Hilbert space for each throat, much as in \cite{McNamara:2020uza}.

Of course, one of the main reasons to seek out an AdS/CFT interpretation of our ensemble average system is its potential use in
studying aspects of quantum gravity in anti-de Sitter space. Along these lines, it is also natural to ask about whether the approximation is reliable enough to provide access to the microscopic details of a black hole. To arrange this, we follow the procedure in \cite{Maldacena:1997re, Witten:1998zw} and consider the CFT on the background $S^1 \times S^{D - 1}$. The size of the thermal circle sets a corresponding temperature (and thus size) for an AdS--Schwarzschild geometry, but one can of course consider more elaborate configurations with various chemical potentials switched on. Now, suppose we are interested in probing the $n_\text{micro}$ microstates of this black hole, perhaps as captured by Hawking radiation quanta. To construct a quantity such as the Page curve, a boundary observer will need to sample order $n^2_\text{micro}$ times \cite{Page:1993wv}. However, if $n_\text{micro}$ exceeds $\sqrt{n_*} \sim \varepsilon^{-1}$ of \cref{eq:nstar}, then the ensemble interpretation becomes problematic. In light of this, it is unclear (at least to us) how we can use this setup to learn about the Page curve of a macroscopic black hole.

It is what it is.

\section{Examples} \label{sec:EXAMPLES}

Having demonstrated the main ideas behind ensemble averaging in the context of string theory, we now turn to explicit examples.
Our aim here is not to be exhaustive, but rather to showcase a few different methods, including their advantages and disadvantages.

For starters, we will focus on supersymmetric quantum field theories since these are the ones over which we have maximal control. We shall also primarily focus on superconformal field theories (SCFTs), since these have a chance (in a suitable large-$N$ limit) of having a holographic dual. Within this setting, we identify two general methods for building an ensemble, one that we refer to as various brane box models (and their dual incarnations), and another based on building up new compactification geometries by fibering an existing geometry over a non-compact subvariety of its moduli space.

One way to build a large class of QFTs is with D-branes that end on NS5-branes, so-called ``brane box'' models. Arranging these NS5-branes in various repeating patterns such as $d$-dimensional generalizations of a cube, we can produce a rich set of possible SCFTs. For our purposes, the important point is that for $D \le 4$ systems, the relative positions between the NS5-branes descend to non-normalizable parameters of the resulting quantum field theory. We get an ensemble average by repeating this construction in some of the directions transverse to the original $d$ dimensions used to make a single instance of the SCFT. For example, if we attempt to engineer a $D$-dimensional QFT using spacetime-filling branes in the geometry $\bR^{D - 1, 1} \times \bR^d \times Y_{10 - D - d}$, then the entire configuration sits at a single point of the transverse $Y$ geometry, which in many cases of interest is just $\bR^{10 - D - d}$. Moving to a different point of $Y$, we can then arrange for a different choice of couplings. By a chain of T-dualities, these constructions can also be related to the worldvolume theory of branes probing singularities, and these in turn can often be generated by appropriate compactifications of 6D SCFTs and little string theories (LSTs). An important feature of this method of construction is that precisely because we can tune the moduli in the transverse $Y$ directions, we can use this to engineer an essentially arbitrary probability distribution over the couplings of the model. The disadvantage of this approach is that in some cases it is difficult to guarantee that we generate a theory with a candidate AdS dual.

Another way to generate examples consists of taking a system of branes wrapped on subspaces of a Calabi--Yau $d$-fold $X_{2 d}$ (a $2 d$-real-dimensional space). More precisely, we assume that the geometry takes the form $\bR^{D - 1, 1} \times Y_{2 m} \times X_{2 d}$, where $Y$ is an $m$-complex-dimensional geometry transverse to the branes. In particular, our branes sit at a particular point of $Y$ and the parameters of the QFT descend from the geometric moduli of $X$. To get an ensemble average, we consider $\cM_X$, the moduli space of $X$. Observe that we can consider the total space $\cX$ as defined by $X \to \cX \to B$, where $B \subset \cM_X$. Cutting out an $m$-complex-dimensional subspace $B$ of $\cM_X$ that has no singular fibers then generates a non-compact Calabi--Yau of real dimension $2 m + 2 d$.\footnote{We thank T.~Pantev for discussions on this point.} We can also entertain more general fibrations, possibly with singular fibers, and we present some explicit examples of precisely this sort in \cref{sec:FIBRATION}. An important advantage of this approach is that such brane constructions in Calabi--Yau compactifications often come with readily defined AdS duals. A drawback of this approach is that it does not, in general, allow us to engineer an arbitrary ensemble average. This is simply because the best we can do is to sweep out a probability distribution with support on an $m$-complex-dimensional subspace of the full moduli space.

In the remainder of this \namecref{sec:EXAMPLES}, we turn to some particular examples, illustrating
the pitfalls, the possibilities, the perils, and the promise of generating ensemble averages.

\subsection{$D = 4$}

We now engineer some ensembles of $D = 4$ SCFTs. We begin by constructing an ensemble for $\cN = 4$ Super Yang--Mills theory, and then turn to examples with lower supersymmetry.

\subsubsection{Warmup: Approximating $\cN = 4$ SYM}

Perhaps the simplest case to consider is that of type IIB string theory on the background $\bR^{3, 1} \times \bC^3$ with a stack of $N_\text{c}$ spacetime-filling D3-branes sitting at a point of $\bC^3$. At low energies, the open string degrees of freedom realize $\cN = 4$ Super Yang--Mills theory with gauge group $\U(N_\text{c})$. The value of the complexified gauge coupling $\tau$ is controlled by the background value of the type IIB axio-dilaton:
\begin{equation}
	\tau = C_{0} + i \exp(-\phi)\,.
\end{equation}
In type IIB supergravity, this is characterized by the extremal brane solution with constant axio-dilaton profile, localized source for the self-dual five-form flux, and metric (see, e.g., \cite{Horowitz:1991cd}):
\begin{equation}
		\diff{s}^2 = H^{-1 / 2} \diff{s}_{\bR^{3, 1}}^2 + H^{1/2} \diff{s}_{\bC^3}^2\,, \quad \text{with } H = 1 + \frac{4 \pi g_\text{s} N_\text{c} \alpha'^2}{r^4}\,,
\end{equation}
where $r$ is the distance from the D3-brane stack. As is well known, in the near-horizon limit, this produces an $\AdS_5 \times S^5$ geometry with $N_\text{c}$ units of self-dual five-form flux threading the two factors \cite{Maldacena:1997re}. The AdS radius and $S^5$ radius $L$ are correlated, and related to the open string parameters as $L^4 = 2 g_\text{YM}^2 N_\text{c} \alpha'^4$. We are, of course, free to consider moving these D3-brane stacks to separate points in $\bC^3$. In all these local sectors, the value of the axio-dilaton is always the same, and we generate a rather trivial probability distribution of values for the axio-dilaton. The supergravity approximation is the same, the only change being the harmonic function $H$, which is now given by
\begin{equation}
	H = 1 + \sum_\stackIndex \frac{4 \pi g_\text{s} N_\text{c} \alpha'^2}{\abs{\V{r} - \V{r}_\stackIndex}^4}\,,
\end{equation}
with $\V{r}_k$ the position of the $k$th stack.

To get a more general class of distributions, we now introduce an additional source as specified by a stack of D7-branes that sits at a point of the middle factor in $\bR^{3,1} \times \bC_\perp \times \bC^2$, and fills the remaining eight directions.
The brane configuration is as follows:
\vspace{2em}
\begin{center}
	\begin{tabularx}{0.5\textwidth}{c|*{10}{>{\centering\arraybackslash $}X<{$}}}
		& \tikzmark{r31L} 0 & 1 & 2 & 3 \tikzmark{r31R} & \tikzmark{cPerpL} 4 & 5 \tikzmark{cPerpR} & \tikzmark{c2L} 6 & 7 & 8 & 9 \tikzmark{c2R} \\ \hline
		D3 & \times & \times & \times & \times & & &        &        &        & \\
		D7 & \times & \times & \times & \times & & & \times & \times & \times & \times
	\end{tabularx}
	\begin{tikzpicture}[overlay, remember picture]
        \draw [decorate,decoration={brace,amplitude=10pt,raise=4pt}] (r31L.west) --node[above=14pt]{$\bR^{3, 1}$} (r31R.east);
        \draw [decorate,decoration={brace,amplitude=10pt,raise=4pt}] (cPerpL.west) --node[above=14pt]{$\bC_\perp$} (cPerpR.east);
        \draw [decorate,decoration={brace,amplitude=10pt,raise=4pt}] (c2L.west) --node[above=14pt]{$\bC^2$} (c2R.east);
    \end{tikzpicture}
\end{center}

Doing so produces a position-dependent profile for the axio-dilaton, but also breaks half the supersymmetry in the system. Indeed,
if we now have a stack of D3-branes located at distinct points of the geometry, then each can experience a different value of the axio-dilaton $\tau_\stackIndex$, and the low-energy effective action on each stack is of the form
\begin{equation}
	S_\stackIndex = S_{\cN = 4}(\tau_\stackIndex) + S_{\cN = 2}\,,
\end{equation}
where the contribution from explicit $\cN = 2$ breaking terms is captured by a collection of higher-dimension operators. The precise form of these contributions can be worked out by noting that this D3/D7 system is just engineering a 4D $\cN = 2$ gauge theory. There are two distance scales that control the strength of these higher-dimension operators. One is the relative separation between the D3-branes,
\begin{equation}
	\Lambda_{\stackIndex_1, \stackIndex_2} \equiv \frac{\dist(\DThree_{\stackIndex_1}, \DThree_{\stackIndex_2})}{\alpha'}\,,
\end{equation}
as measured in the full $\bC^3$ factor transverse to all the D3-branes,
and the other is the relative separation between the D3-branes and the D7-branes,
\begin{equation}
	\Lambda_{\stackIndex, \DSeven} \equiv \frac{\dist(\DThree_\stackIndex, \DSeven)}{\alpha'}\,,
\end{equation}
as measured in the $\bC_\perp$ factor transverse to the D7-branes. Provided we only ask questions at low energies compared with these cutoffs, we get an adequate approximation to ensemble averaging in $\cN = 4$ SYM, but one that has an $\cN = 2$ UV completion.

Let us now turn to the class of ensembles we can actually engineer in this setting. First of all, the whole point of introducing a stack of D7-branes is that we can thus generate a position-dependent axio-dilaton. More broadly, this and more general choices of non-perturbative bound states of 7-branes can be understood in F-theory \cite{Vafa:1996xn, Morrison:1996na, Morrison:1996pp} by considering a non-compact elliptically fibered K3 surface with minimal Weierstrass model:
\begin{equation}
	y^2 = x^3 + f(z) x + g(z)\,,
\end{equation}
where in the present setting, $f(z)$ and $g(z)$ are treated as polynomials in the holomorphic coordinate $z$ of the $\bC_\perp$ factor. The possible values of the axio-dilaton are implicitly encoded in the $\SL(2, \bZ)$-invariant $j$-function:
\begin{equation}
	j = 1728 \frac{4 f^3}{4 f^3 + 27 g^2}\,,
\end{equation}
which has the weak-coupling expansion in $q = \exp(2 \pi i \tau)$ given by
\begin{equation}
	j = q^{-1} + 744 + 196884 q + \dotsb\,.
\end{equation}
For example, $j = \infty$ corresponds to weak coupling at $\tau = i \infty$, and $j = 1728$ corresponds to $\tau = i$, while $j = 0$ corresponds to $\tau = \exp(2 \pi i / 6)$. In the case of a single stack of $M$ D7-branes sitting at $z = 0$, we just have
\begin{equation}
	j_{\DSeven} = z^{-M}\,.
\end{equation}
To get a particular value of the axio-dilaton (or, more precisely, its $j$-invariant), we simply consider D3-branes at the desired value of $z$. Note that this intrinsically comes with some limitations, because to populate the distribution near the $\tau = i \infty$ region of moduli space, we necessarily must move close to the stack of D7-branes, which in turn lowers the UV scale $\Lambda_{\stackIndex, \DSeven}$ in our effective field theory. The other issue we face is how to sequester the D3-brane stacks from one another. This is less problematic, because even if they sit at the same point of $\bC_\perp$, we are free to move them away from each other in the $\bC^2$ factor. We thus conclude that the UV cutoff is set by $\Lambda_{\stackIndex, \DSeven}$, and this in turn depends on what sort of distribution we wish to engineer.

Consider next the supergravity background generated by our D3/D7 system. Since we are dealing with D3-brane probes of an F-theory geometry, the main change is that the metric on the $\bC_\perp$ factor is controlled by that of the non-compact elliptically fibered K3 space $\KThree \to \bC_\perp$. In particular, we observe that in the near-horizon limit for each local stack, we indeed get a collection of individual AdS throats, but of different sizes as dictated by the local profile of the $\tau_\stackIndex$. Note also that the subleading $\cN = 2$ breaking terms amount to a deviation away from a pure $\AdS_5 \times S^5$ geometry.

Precisely because the value of the bulk cosmological constant is different for each local sector, there is no sense in which we can give a holographic interpretation in terms of a single AdS geometry. This is in accord with the fact that consistency of the MM proposal requires the baby universe Hilbert space interpretation to be trivial (i.e., one-dimensional) in this special case \cite{Marolf:2020xie, McNamara:2020uza}.

\subsubsection{Quiver Gauge Theory Ensemble}

\begin{figure}[t!]
	\centering

	\includegraphics[width=4cm]{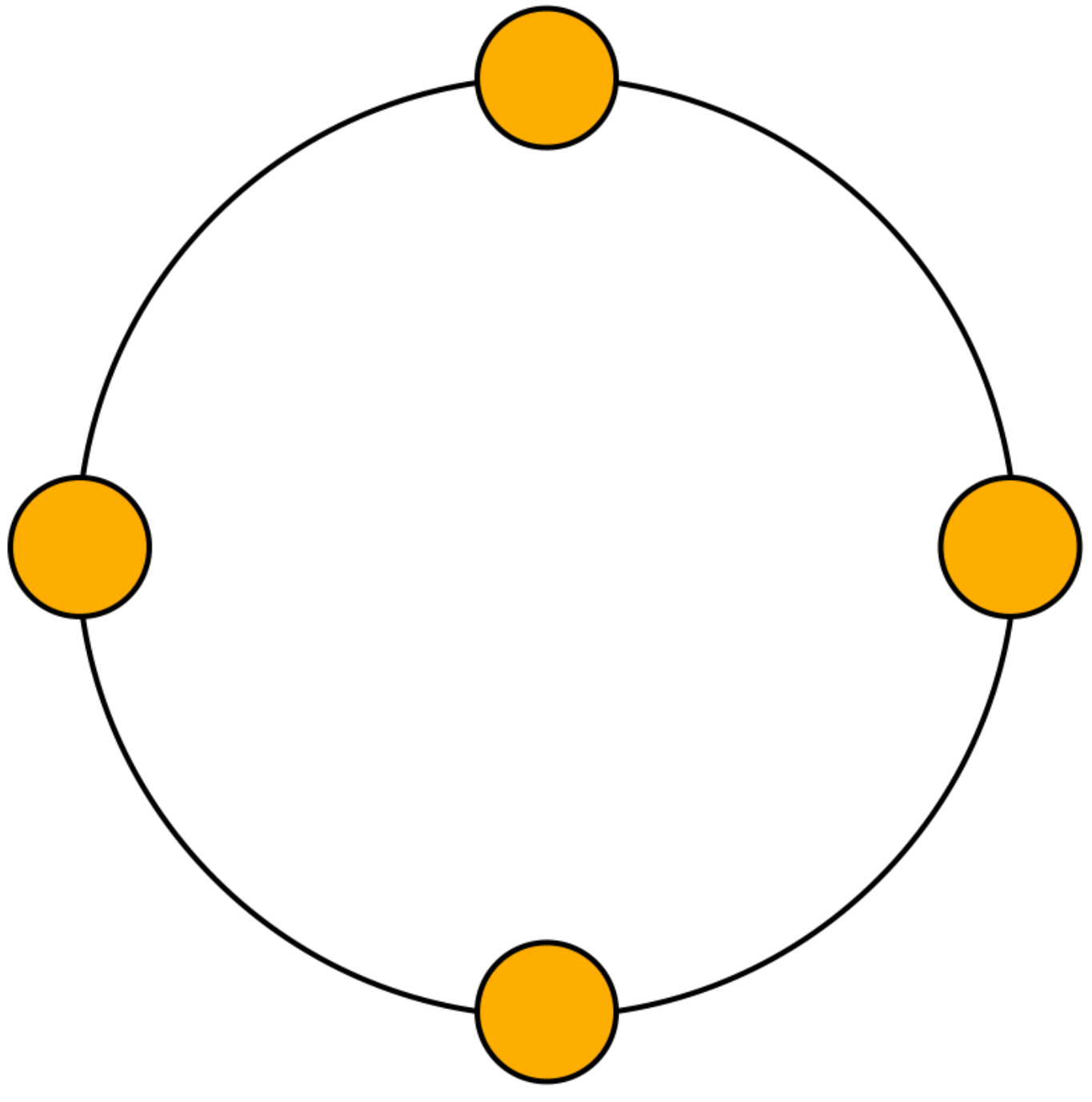}

	\caption{Quiver diagram of the 4D $\cN = 2$ SCFT obtained from $N_\text{c}$ D3-branes probing $\bC^2 / \bZ_M$, for $M = 4$.
Each node represents an $\SU(N_\text{c})$ gauge group, and links between them denote bifundamental hypermultiplets.}
	\label{fig:N=2 quiver}
\end{figure}

We now proceed to engineer an ensemble average of quiver gauge theories that enjoys a holographic dual. The model we consider consists of $N_\text{c}$ D3-branes probing the orbifold singularity $\bC^2 / \bZ_M$ with group action $(u, v) \mapsto (\xi u, \xi^{-1} v)$ on the holomorphic coordinates, where $\xi$ is a primitive $M$th root of unity. As is well known from \cite{Douglas:1996sw, Lawrence:1998ja}, the worldvolume theory for this model results in a 4D $\cN = 2$ SCFT described by a quiver gauge theory with gauge groups arranged in a circular ring, joined by hypermultiplets in bifundamental representations (see \cref{fig:N=2 quiver}). The model also comes with a collection of marginal parameters, captured by the holomorphic couplings $(\tau^{(1)}, \dotsc, \tau^{(M)}) \equiv \V{\tau}$. These are encoded in the choice of closed string moduli.

To better understand the sense in which these closed string moduli are tunable, it is helpful to consider some dual realizations of the same low-energy effective field theory. One way to proceed is to observe that there is a 6D little string theory obtained via F-theory from a configuration of collapsing $-2$-curves arranged in a circular ring (see, e.g., \cite{Bhardwaj:2015oru}). Wrapping $N_\text{c}$ D7-branes on each $-2$-curve (a Kodaira $I_{N_\text{c}}$ fiber) then results in a 6D quiver gauge theory. Compactifying on a further $T^2$ then produces the desired 4D $\cN = 2$ SCFT.
In this realization, the gauge couplings descend from the complexified K{\"a}hler volume of $T^2 \times \Sigma_m$, for $m = 1, \dotsc, M$ labelling the different $-2$ curves. In particular, each of these is a tunable complexified gauge coupling in the 4D field theory.

The tensor branch of the 6D little string theory can also be realized from a configuration of $M$ NS5-branes arranged in a circular ring with $N_\text{c}$ D6-branes suspended in between each neighboring pair. In this picture, the relative distance between each NS5-brane sets the value of 6D gauge coupling, and further compactification on a $T^2$ again results in the same 4D gauge theory.

\begin{figure}[t!]
	\centering

	\includegraphics[width=11cm]{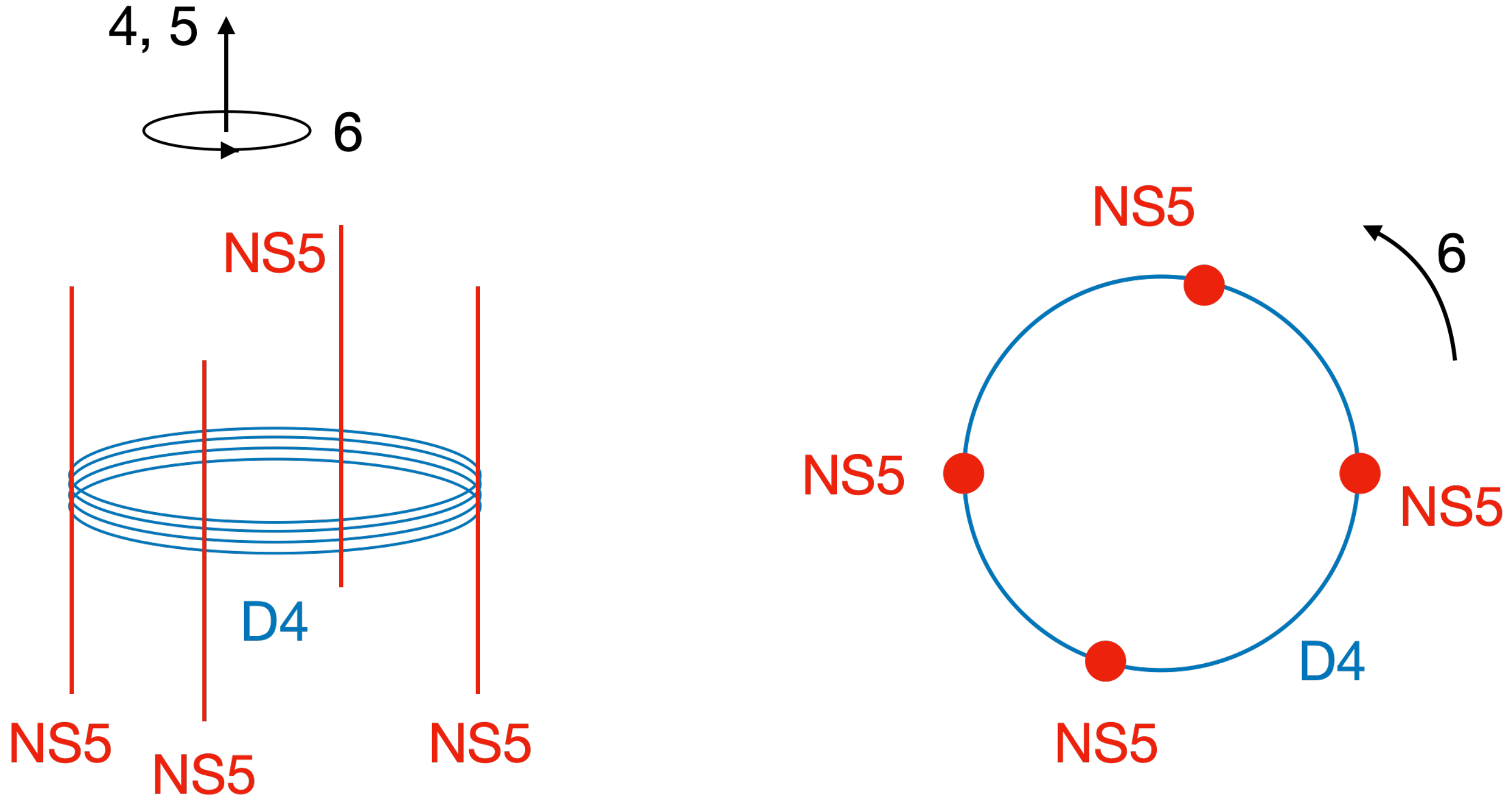}

	\caption{D4/NS5 system leading to the 4D quiver gauge theory. Each D4-brane segment corresponds to a gauge group, whose coupling is set by the relative distance in the 6 direction between the two boundary NS5-branes as $\frac{1}{g_i^2}=\frac{x_{i + 1}^6 - x_i^6}{g_\text{s} \sqrt{\alpha'}}$. The effective 4D SCFT on the D4-brane worldvolume is the same quiver gauge theory shown in \cref{fig:N=2 quiver}.}
	\label{fig:D4-NS5}
\end{figure}

To be more explicit, let us consider T-dualizing the $T^2$ wrapped by D6-branes; the 4D gauge theory is then realized alternatively by the following D4/NS5 system:
\begin{equation*}
	\begin{array}{c|*{10}{c}}
		          & 0      & 1      & 2      & 3      & 4 & 5 & 6      & 7      & 8      & 9 \\ \hline
		N_\text{c} \text{ D4s} & \times & \times & \times & \times &   &   &  \times      &        &        & \\
		M \text{ NS5s} & \times & \times & \times & \times &  \times & \times  &  &  &  &
	\end{array}
\end{equation*}
See \cref{fig:D4-NS5} for an illustration of this brane configuration in the 4, 5, and 6 directions. Observe also that a T-duality on direction 6 directly connects this construction to that of $N_\text{c}$ D3-branes probing the same $\bC^2 / \bZ_M$ singularity. In all these cases, then, we have a geometric characterization of the resulting moduli.

Let us now turn to the construction of an ensemble average. Returning to the picture of D3-branes probing an orbifold, observe that we can generate a large ensemble of local singularities via the hypersurface in $\bC^3$ given by:
\begin{equation}
	y^2 = x^2 + \prod_\stackIndex (z - z_\stackIndex)^{M_\stackIndex}\,.
\end{equation}
Near $x = y = (z - z_\stackIndex) = 0$, this generates a $\bC^2 / \bZ_{M_\stackIndex}$ singularity, so if we take $M_\stackIndex = M$ for all $\stackIndex$, and place $N_\text{c}$ D3-branes near each singularity, we produce an ensemble of such systems. Observe that because the closed string moduli of each singularity are decoupled, we can arrange a nearly arbitrary ensemble. The same is clear from working with any of the other local pictures. Not also that there is a natural UV completion available in terms of a stack of $\numStacks N_\text{c}$ D3-branes probing a $\bC^2 / \bZ_{\numStacks M}$ singularity.

In the large-$N_\text{c}$ limit, we can also arrange for a holographic dual description. To get the same size AdS, we just require that the value of the string theory dilaton (in the D3-brane probe picture) is the same for each stack. In the quiver gauge theory, this is controlled by the particular combination of parameters
\begin{equation}
	\tau_\text{IIB} = \tau^{(1)} + \dotsb + \tau^{(M)}
\end{equation}
at each local model, which we take to be equal for each value of $\stackIndex = 1, \dotsc, \numStacks$. In this description, the remaining $M - 1$ degrees of freedom are tunable marginal couplings over which our ensemble runs.

Clearly, one can consider far more elaborate examples, but the essential point is that at least for $D = 4$ SCFTs, we can engineer a wide variety of ensemble averages.

\subsection{$D = 3$}

Let us briefly comment on the construction of $D = 3$ SCFTs, and issues with engineering ensemble averages in this setting. First of all, we can just take a $D = 4$ SCFT and compactify it on a circle. In many cases, this can also result in a 3D SCFT. From the perspective of the examples just considered, we could alternatively start from a 6D LST or 6D SCFT and compactify it on a three-manifold with negative sectional curvature. The primary challenge from this perspective is how to build a class of compactification geometries where we can vary the marginal couplings.
\footnote{A general class of 3D SCFTs can be obtained by compactification of 6D $\cN = (2,0)$ SCFTs on a three-manifold, which are known as the $T[M_3]$ theories (see, e.g., \cite{Dimofte:2011ju,Gadde:2013sca}). When $M_3 = \Sigma \times S^1$, where $\Sigma$ is a Riemann surface, $T[M_3]$ admits marginal deformations. However, the holographic dual of $T[M_3]$ in this case is not $\AdS_4$ gravity, but rather a ``gravitational domain wall" separating two AdS regions (see, e.g., \cite{Gukov:2015qea}). Even if one focuses on the CFT side, the top-down approach to building an ensemble average of these 3D SCFTs is not clear to us. We thank S.~Gukov for discussions on this point.}

To this general point, one might attempt to realize examples using M-theory on an eight-manifold with a large number of singularities. The theory of $N_\text{c}$ M2-branes probing any one singularity will produce an $\AdS_4 \times X_7$ background, and in principle the geometric moduli of $X_7$ can be traced to corresponding deformations in the 3D SCFT. One can consider, for instance, an ensemble average over the Chern--Simons levels of the ABJ(M) model \cite{Aharony:2008ug, Aharony:2008gk} by introducing $\numStacks$ copies of $N_\text{c}$ M2-branes probing $\bC^4 / \bZ_{M_\stackIndex}$ singularities. The difficulty we now face is that the radius of the dual $\AdS_4$ scales as $L^2 \propto N_\text{c} / M_\stackIndex$ with other parameters fixed. On the other hand, if we attempt to average over the Chern--Simons couplings, we must then simultaneously adjust the value of $N_\text{c}$. It would be interesting to build an explicit ensemble average in this case, but this would seem to require first obtaining a better understanding of how the geometry of a given background descends to marginal parameters of
the field theory.\footnote{See references \cite{Gauntlett:2005jb, Imeroni:2008cr, Bobev:2021gza, Bobev:2021yya} for some examples of how these marginal couplings are realized in various $\AdS_4$ gravity duals. However, the geometric or brane realization counterparts of these deformations are not yet known, so it is unclear to us how to build an ensemble with random couplings from a top-down approach.
We thank N.~Bobev and A.~Tomasiello for helpful correspondence on this point.}

\subsection{$D = 2$}

Let us now turn to the construction of $D = 2$ SCFTs, and, when appropriate, their $\AdS_3$ duals.
We begin with an example of a brane box model where we have a great deal of freedom in generating
an ensemble average. The downside, however, is that the existence of an $\AdS_3$ dual is not always apparent.
The other general method we consider yields an $\AdS_3 \times S^3 \times X_4$ dual, but at the expense of a less-flexible class of possible probability distributions.

\subsubsection{Brane Box Examples}

Similarly to 4D SCFTs, in addition to branes probing geometries, one can also engineer 2D theories using intersecting branes. Let us now consider 2D $\cN = (0, 2)$ theories engineered by the brane box model \cite{Garcia-Compean:1998sla,Franco:2015tya}.

\begin{figure}[t!]
	\centering

	\includegraphics[width=10cm]{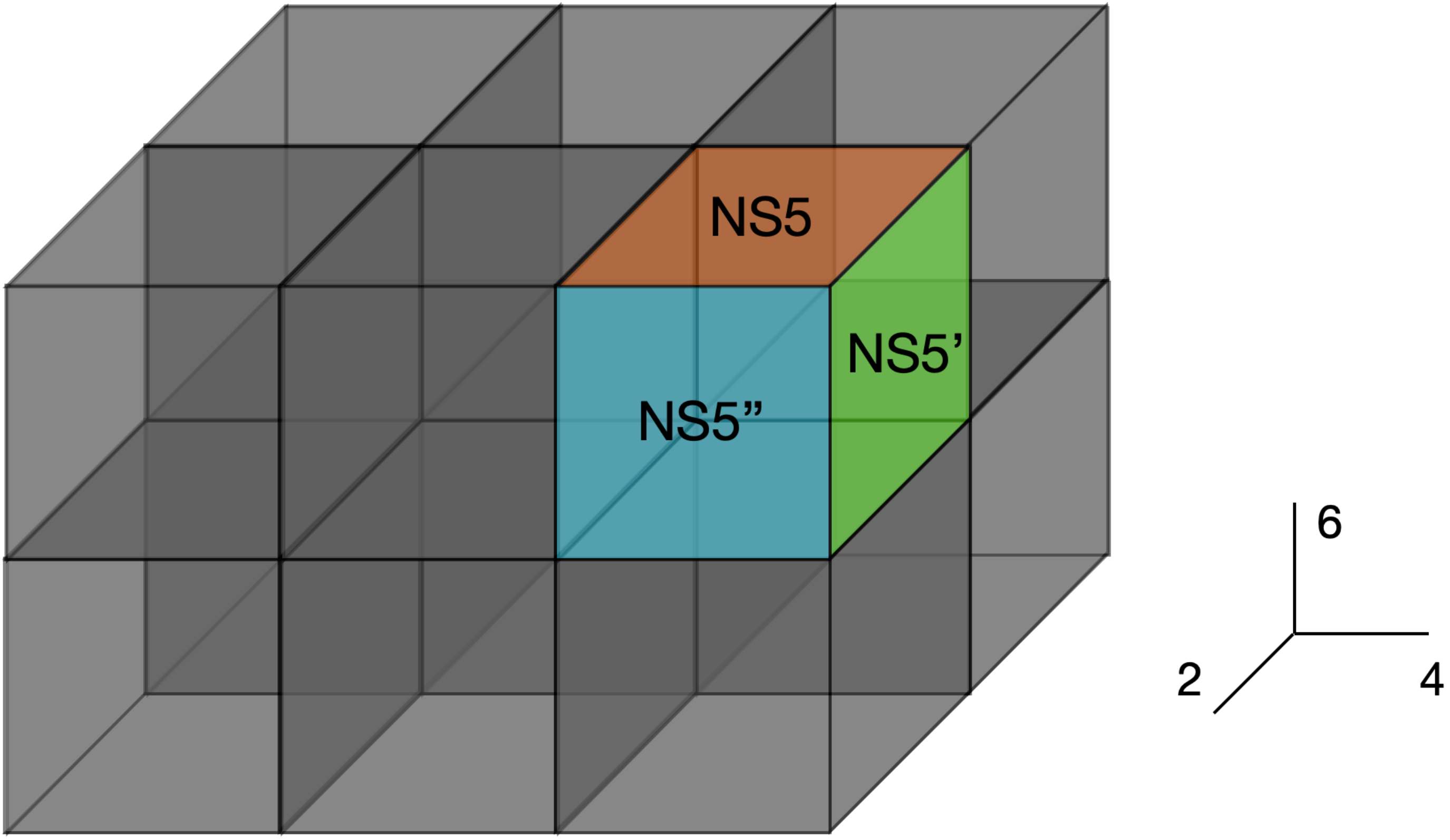}

	\caption{Brane box model on $T^3$ with $\numNSFives = 2, \numNSFives' = 3$ and $\numNSFives'' = 2$. Every box bounded by NS5-, NS5$'$-, and NS5$''$-branes is filled by $M$ D4-branes, corresponding to one $\U(M)$ gauge group in the 2D effective field theory.}
	\label{fig:brane box}
\end{figure}

The brane configuration consists of $M$ D4-branes, $\numNSFives$ NS5-branes, $\numNSFives'$ NS5$'$-branes, and $\numNSFives''$ NS5$''$-branes, as follows:
\begin{equation*}
	\begin{array}{c|*{10}{c}}
		             & 0      & 1      & 2      & 3      & 4      & 5         & 6         & 7         & 8 & 9 \\ \hline
		\text{D4}    & \times & \times & \times &        & \times &           & \times    &           &   & \\
		\text{NS5}   & \times & \times & \times & \times & \times & \times    &           &           &   & \\
		\text{NS5}'  & \times & \times & \times & \times &        &           & \times    & \times    &   & \\
		\text{NS5}'' & \times & \times &        &        & \times & \times    & \times    & \times    &   &
	\end{array}
\end{equation*}
where all branes sit on the same position in the $8, 9$-plane.  The D4-branes are finite in the 2, 4, and 6 directions as a $T^3$, and bounded by three types of NS5-branes. For each pair of parallel NS5-branes, the brane configuration in directions 2, 4, and 6 is a 3D box filled by D4-branes. With $\numNSFives$ NS5-branes, $\numNSFives'$ NS5$'$-branes, and $\numNSFives''$ NS5$''$-branes, we have a 3D grid containing $\numNSFives \cdot \numNSFives' \cdot \numNSFives''$ boxes, as illustrated in \cref{fig:brane box}.

The effective field theory on D4-branes in the non-compact 0 and 1 directions is a 2D $\cN = (0, 2)$ gauge theory with $\U(M)^{\numNSFives \cdot \numNSFives' \cdot \numNSFives''}$ gauge symmetry. Each $\U(M)$ gauge group corresponds to one brane box on the $T^3$ comprising the 2, 4, and 6 directions. Performing T-duality on this $T^3$, the brane configuration becomes D1-branes probing a $\bC^4 / (\bZ_\numNSFives \times \bZ_{\numNSFives'} \times \bZ_{\numNSFives''})$ singularity, with orbifold group action:
\begin{equation}
\begin{aligned}
	(z_1, z_2, z_3, z_4) &\to (z_1, z_2, e^{\frac{2 \pi i}{\numNSFives}} z_3, e^{\frac{2 \pi i}{\numNSFives}} z_4)\,, \\
	(z_1, z_2, z_3, z_4) &\to (z_1, e^{\frac{2 \pi i}{\numNSFives'}} z_2, z_3, e^{\frac{2 \pi i}{\numNSFives'}} z_4)\,, \\
	(z_1, z_2, z_3, z_4) &\to (e^{\frac{2 \pi i}{\numNSFives''}} z_1, z_2, z_3, e^{\frac{2 \pi i}{\numNSFives''}} z_4)\,,
\end{aligned}
\end{equation}
which indeed preserves $\cN = (0, 2)$ supersymmetry on the 2D worldvolume of the D1-branes.

Due to the non-vanishing elliptic genera of these 2D $\cN = (0, 2)$ theories computed in \cite{Franco:2017cjj}, it is natural to conjecture that these theories flow to SCFTs in the IR.\footnote{To our knowledge, the explicit SCFTs at the IR fixed point and the AdS gravity duals for these $\cN = (0,2)$ theories have not yet been constructed, which by itself is a problem deserving further investigation.} Here, we will assume the existence of the SCFT in the IR and build the ensemble average.\footnote{We thank N.~Benjamin, S.~Franco, and S.~Gukov for helpful discussions on this point.}

One of the marginal couplings in 2D is the Fayet--Iliopoulos (FI) term for the $\U(1)$ factor of each $\U(N)$ gauge group. The corresponding FI parameters are encoded in the positions of the NS5-branes in several directions. Namely, for a given $\U(M)$ gauge group associated with the $(\numNSFives, \numNSFives', \numNSFives'')$-th brane box, its FI parameter is given in terms of the separations of the NS5-, NS5$'$-, and NS5$''$-branes in directions 7, 5, and 3, respectively (see, e.g., \cite{Garcia-Compean:1998sla}):
\begin{equation}
	r_{\numNSFives, \numNSFives', \numNSFives''} = \frac{(x^7_{\numNSFives + 1} - x^7_\numNSFives) + (x^5_{\numNSFives' + 1} - x^5_{\numNSFives'}) + (x^3_{\numNSFives'' + 1} - x^3_{\numNSFives''})}{\sqrt{\alpha'}}.
\end{equation}

We can build an ensemble of brane box models by repeating this construction at different points in the $8, 9$-plane. Indeed, tuning the relative positions of the NS5-, NS5$'$-, and NS5$''$-branes in each instance of the construction, we can engineer an ensemble of 2D $\cN = (0, 2)$ SCFTs in which the FI-parameters are drawn from a probability distribution.

A drawback of this approach is that although this provides us with a way to generate ensemble averaging in 2D SCFTs, the existence of a putative AdS dual is somewhat unclear. We now turn to examples for which we understand the holographic dual more clearly.

\subsubsection{Holographic Example}

To generate some examples with a holographic dual, we consider type IIB string theory on the background $\bR_\text{time} \times \bC^2 \times S^1 \times \KThree$, with $N_5$ D5-branes wrapped on $S^1 \times \KThree$ and $N_1$ D1-branes wrapped on the $S^1$ factor. We keep the D1/D5 system coincident at the same point of $\bC^2$. This engineers a 2D SCFT with $\cN = (4, 4)$ supersymmetry on the spacetime $\bR_\text{time} \times S^1$. The brane configuration is as follows:
\vspace{2em}
\begin{center}
	\begin{tabularx}{0.6\textwidth}{c|*{10}{>{\centering\arraybackslash $}X<{$}}}
		& \tikzmark{rTimeL} 0 \tikzmark{rTimeR} & \tikzmark{s1L} 1 \tikzmark{s1R} & \tikzmark{c2L} 2 & 3 & 4 & 5 \tikzmark{c2R} & \tikzmark{k3L} 6 & 7 & 8 & 9 \tikzmark{k3R} \\ \hline
		D1 & \times & \times & & & & &        &        &        & \\
		D5 & \times & \times & & & & & \times & \times & \times & \times
	\end{tabularx}
	\begin{tikzpicture}[overlay, remember picture]
        \draw [decorate,decoration={brace,amplitude=10pt,raise=4pt}] (rTimeL.west) --node[above=14pt]{$\bR_\text{time}$} (rTimeR.east);
        \draw [decorate,decoration={brace,amplitude=10pt,raise=4pt}] (s1L.west) --node[above=14pt]{$S^1$} (s1R.east);
        \draw [decorate,decoration={brace,amplitude=10pt,raise=4pt}] (c2L.west) --node[above=14pt]{$\bC^2$} (c2R.east);
        \draw [decorate,decoration={brace,amplitude=10pt,raise=4pt}] (k3L.west) --node[above=14pt]{$\KThree$} (k3R.east);
    \end{tikzpicture}
\end{center}

The corresponding supergravity solution describes an extremal black string, and in the near-horizon limit, it is given by the geometry $\AdS_3 \times S^3 \times \KThree$, which figures prominently in the $\AdS_3 / \CFT_2$ correspondence \cite{Maldacena:1997re} (see also \cite{Brown:1986nw}). The size of the AdS radius in Planck units is proportional to $N_1 N_5$, which is in turn proportional to the central charge of the 2D SCFT $c = 6 N_1 N_5$. For our purposes, the important point is that this 2D SCFT comes with a set of marginal couplings controlled by the metric moduli of the K3 surface, and geometrically, these can be understood (via duality) as Narain moduli of the sort that appear in \cite{Maloney:2020nni,Afkhami-Jeddi:2020ezh,Benjamin:2021wzr}.

To generate an ensemble, we begin by observing that $\bC^2 = \bC_\text{seq} \times \bC_\text{base}$ can be written as a product of two complex lines. We shall use the $\bC_\text{seq}$ factor to sequester the different QFTs from one another, and will use the $\bC_\text{base}$ factor to build a K3-fibered Calabi--Yau threefold $X_3 \to \bC_\text{base}$ with fiber given by our K3 surface. Populating a different number of local sectors at each value of the moduli, we can again construct a probability distribution in a one-parameter subspace of the K3 moduli space. Let us also note that, much as in our discussion of the D3/D7 system, approaching a singular point in the moduli space of K3s limits the regime of validity of our approximation.

\subsection{$D = 1$}

Finally, let us turn to the case of one-dimensional quantum mechanical systems and their approximate $\AdS_2$ duals. Along these lines, we observe that type IIA strings on $\bR_\text{time} \times \bR^3 \times X$, with $X$ a Calabi--Yau threefold, produces a 4D $\cN = 2$ supergravity background. Wrapping D0-, D2-, D4-, and D6-branes on holomorphic cycles of $X$ can, for suitable charges, result in a 4D black hole solution with near-horizon limit $\AdS_2 \times S^2$, with radius set by the background value of the charges (see, e.g., \cite{Simons:2004nm}). The brane configuration is shown below:
\vspace{2em}
\begin{center}
	\begin{tabularx}{0.6\textwidth}{c|*{10}{>{\centering\arraybackslash $}X<{$}}}
		& \tikzmark{rTimeL} 0 \tikzmark{rTimeR} & \tikzmark{r3L} 1 & 2 & 3 \tikzmark{r3R} & \tikzmark{xL} 4 & 5 & 6 & 7 & 8 & 9 \tikzmark{xR} \\ \hline
		D0 & \times & & & &        &        &        &        &                   & \\
		D2 & \times & & & & [      & \multicolumn{4}{c}{\text{2 of 6 directions}} & ] \\
		D4 & \times & & & & [      & \multicolumn{4}{c}{\text{4 of 6 directions}} & ] \\
		D6 & \times & & & & \times & \times & \times & \times & \times            & \times
	\end{tabularx}
	\begin{tikzpicture}[overlay, remember picture]
        \draw [decorate,decoration={brace,amplitude=10pt,raise=4pt}] (rTimeL.west) --node[above=14pt]{$\bR_\text{time}$} (rTimeR.east);
        \draw [decorate,decoration={brace,amplitude=10pt,raise=4pt}] (r3L.west) --node[above=14pt]{$\bR^3$} (r3R.east);
        \draw [decorate,decoration={brace,amplitude=10pt,raise=4pt}] (xL.west) --node[above=14pt]{$X$} (xR.east);
    \end{tikzpicture}
\end{center}
Here, the bracketed directions denote the subset of directions containing the cycles wrapped by D2- and D4-branes, but do not specify the exact directions those cycles fill out. In the 4D theory, the D0- and D2-brane charges give rise to an electric charge vector $\V{Q}$, while the D4- and D6-brane charges give rise to a magnetic charge vector $\V{P}$. In general, it is a challenging problem to construct multi-center black hole solutions, but in the special case where all the centers have the \emph{same} charge $(\V{Q}_\stackIndex, \V{P}_\stackIndex) = (\V{Q}, \V{P})$, the configurations are mutually BPS, and in particular can be separated from one another in arbitrary directions of the $\bR^{3}$ factor. For our purposes, we shall find it convenient to write $\bR^3 = \bR_\text{seq} \times \bC_\text{base}$.

The worldvolume theory of this configuration is also challenging to describe, but at least at weak string coupling (away from the supergravity limit), there is a corresponding quiver quantum mechanics description available (see, e.g., \cite{Denef:2002ru}).
In the type IIA description, the superpotential couplings of this model are controlled by the complex structure moduli of $X$, and in the mirror type IIB description, they are captured by the K{\"a}hler moduli. An important open problem is to explain the sense in which this quiver quantum mechanics actually ``flows'' to a 1D SCFT.\footnote{Of course, in one dimension we face the issue that the CFT condition of a traceless stress tensor would appear to trivialize the theory altogether. So, we must already be prepared to work in terms of an approximate notion of conformal invariance where we introduce an explicit IR cutoff from the start, much as in \cite{Maldacena:2016hyu}.}

To generate an ensemble average in the 1D system, we return to the decomposition $\bR^3 = \bR_\text{seq} \times \bC_\text{base}$. Since we can retain a supersymmetric multi-center configuration no matter how we separate the branes in the $\bR^3$ direction, we use the $\bR_\text{seq}$ factor to sequester the black holes, and treat $\bC_\text{base}$ as the base of a non-compact Calabi--Yau fourfold $X_4 \to \bC_\text{base}$, with fiber given by $X$ at a specified value of the moduli. Much as in our 2D holographic example, this provides us with an ensemble average over a one-parameter subspace of the moduli space.\footnote{The disorder averaged quiver quantum mechanics has been studied in \cite{Anninos:2016szt}, where the probability distribution is chosen to be Gaussian. Our construction in this section can be regarded as why/how the ensemble arise in the quiver quantum mechanics.}

An important subtlety with this construction is that as we vary the Calabi--Yau moduli, we may be forced to deal with a further fragmentation of a single center solution into a multi-center solution (the supergravity analog of wall-crossing), as studied for example in \cite{Denef:2001xn,Denef:2007vg}. For our present purposes, this introduces some additional complications into the $\CFT_1$ description and any putative $\AdS_2$ dual. One can view this as either imposing additional restrictions on what sorts of distributions we can engineer, or alternatively, as an opportunity to come to grips with multiple large $\AdS_2$ throats right from the start.\footnote{We thank N.~Bobev for comments.} As an additional comment, let us note that if we move away from the truncated subsector of operators, we get a large number of $\AdS_2 \times S^2$ throats, much as in the stringy baby universe construction of reference \cite{Dijkgraaf:2005bp}.

\subsection{$D>4$}

For SCFTs in $D = 5$ or $6$,\footnote{Note that $D = 6$ is the highest spacetime dimension where SCFTs exist \cite{Nahm:1977tg}.} it was demonstrated in \cite{Louis:2015mka,Cordova:2016xhm} that there is no supersymmetry-preserving marginal deformation. It would be interesting to address this issue in the context of the MM picture, since it implicitly imposes restrictions on the sort of boundary conditions for the bulk gravitational path integral.

\section{Discussion}\label{sec:DISC}

In this paper, we have presented a string-based method for engineering ensemble averaged QFTs. The main idea in our construction is to first build multiple copies of the same QFT, but in which the non-normalizable modes of the background geometry vary in the transverse directions. Truncating to the subsector of operators that are distributed over all the local sectors, we have shown that this can be used to build up an approximation to ensemble averaging with a UV completion. The construction can also be used to realize models with an approximate AdS dual. There is an intrinsic regime of validity that can be detected just by sampling macroscopic objects a large number of times. This is important because it cuts to the core issue of whether holography and effective field theory can actually be combined without reference to any putative UV completion. Our (admittedly contrived) construction serves as a counterexample to the claim that a UV completion is ``not necessary''.

In light of these considerations, it would be interesting to revisit the computation of a Page curve for a macroscopic black hole (see, e.g., \cite{Penington:2019npb, Almheiri:2019psf, Almheiri:2019hni}). For some recent critiques of these calculations, see, e.g., \cite{Geng:2020qvw,Geng:2021hlu,Raju:2021lwh}.

It is also natural to ask whether we could directly couple our ensemble of QFTs back to gravity. At least in the context of string theory, this is deeply problematic because in all known constructions, there are sharp upper bounds on the total number of extra sectors we can introduce. Moreover, once we recouple to gravity, all the moduli will once again become dynamical and must be stabilized in some way.

Aside from these intrinsic limitations of our method of construction, it is also interesting to ask about the extent to which we can generate more general classes of probability distributions. For example, in our examples based on fibering a Calabi--Yau manifold over its moduli space, we are automatically restricted in the sorts of distributions we can build. It would be helpful to pinpoint whether there is a deeper reason for such constraints.

\newpage

\section*{Acknowledgments}

We thank V.~Balasubramanian, N.~Benjamin, N.~Bobev, S.~Franco,
S.~Gukov, H.~Maxfield, J.~McNamara, T.~Pantev, D.~Pei, M.~Porrati,
A.~Tomasiello, and C.~Vafa for helpful discussions and correspondence.
We thank N.~Bobev, J.~Magan, and E.~Shaghoulian for helpful comments
on an earlier draft. JJH and XY thank the 2021 Simons
summer workshop at the Simons Center for Geometry and Physics for
kind hospitality during part of this work. XY thanks the UPenn theory group
for hospitality during this work. XY also thanks the Caltech Particle Theory Group
for hospitality during part of this work. The work of JJH and APT is supported by the DOE (HEP)
Award DE-SC0013528.

\appendix

\section{Calabi--Yau Moduli Space Fibration}\label{sec:FIBRATION}

In this \namecref{sec:FIBRATION}, we discuss how to produce a Calabi--Yau $(d + m)$-fold as a fibration of a Calabi--Yau $d$-fold $X$ over some $m$-dimensional subvariety $B \subset \cM_X$ of its moduli space $\cM_X$.

As discussed in the main text, adjunction tells us that we can produce a Calabi--Yau total space by fibering a Calabi--Yau over a Calabi--Yau subvariety of its own moduli space.\footnote{For our purposes, we actually consider the pullback of such a fibration by a bijection from a subspace of spacetime onto this subvariety of the moduli space.} However, this story is complicated by singularities in the fiber, and so we will also consider here some more specific examples.

Explicitly, let us consider the case that the $d$-fold is defined as the vanishing locus of a homogeneous degree-$(d + 2)$ polynomial
	\begin{equation}
		p_{d + 2} = \sum_{i_0 + \dotsb + i_{d + 1} = d + 2} a_{i_0, \dotsc, i_{d + 1}} x_0^{i_0} \dotsm x_{d + 1}^{i_{d + 1}}
	\end{equation}
in $\bP^{d + 1}$. To fiber this over an $m$-complex-dimensional base $B$, we promote $a_{i_0, \dotsc, i_{d + 1}}$ and $x_j$ to sections of line bundles:
	\begin{equation}
		\begin{aligned}
			x_j &\in \Gamma(Y, \pi^*(\cL_j) \otimes \cO_Y(1))\,, \\
			a_{i_0, \dotsc, i_{d + 1}} &\in \Gamma(B, \cL \otimes \bigotimes_j \cL_j^{-i_j})\,,
		\end{aligned}
	\end{equation}
where $Y = \bP^{d + 1}(\cL_0 \oplus \dotsb \oplus \cL_{d + 1})$ is the ambient space, $\pi$ is the projection map of the fibration $\pi\colon Y \to B$, and $\cL, \cL_j$ are line bundles over $B$. The total space $\cX$ of the fibration is then a hypersurface $p_{d + 2} = 0$ in the bundle $Y$ over $B$. The total Chern class of this bundle is
	\begin{equation}
		c(Y) = c(B) \prod_j (1 + c_1(\cL_j) + c_1(\cO_Y(1)))\,.
	\end{equation}
By adjunction, we can then compute the total Chern class of $\cX$ as
	\begin{equation}
		c(\cX) = \frac{c(Y)}{1 + c_1(\cL) + (d + 2) c_1(\cO_Y(1))}\,,
	\end{equation}
giving the first Chern class
	\begin{equation}
		c_1(\cX) = c_1(B) + \sum_j c_1(\cL_j) - c_1(\cL)\,.
	\end{equation}
Thus, the Calabi--Yau condition is simply
	\begin{equation}
		c_1(\cL) = c_1(B) + \sum_j c_1(\cL_j)\,.
	\end{equation}

As an explicit example, consider the case of the quintic threefold in $\bP^4$ given by the hypersurface equation
	\begin{equation}
		\sum_{i = 0}^4 \mu x_i^5 - 5 \psi x_0 x_1 x_2 x_3 x_4 = 0\,,
	\end{equation}
with $[x_0 : x_1 : x_2 : x_3 : x_4]$ homogeneous coordinates on $\bP^4$. This is a generalization of the familiar Dwork family of quintics, which corresponds to $\mu = 1$. In this case, by homogeneity we have
	\begin{equation}
		\begin{aligned}
			x_j &\in \Gamma(Y, \pi^*(\cL_x) \otimes \cO_Y(1))\,, \\
			\mu, \psi &\in \Gamma(B, \cL \otimes \cL_x^{-5})\,.
		\end{aligned}
	\end{equation}
Thus, the Calabi--Yau condition becomes
	\begin{equation}
		c_1(\cL) = c_1(B) + 5 c_1(\cL_x)\,.
	\end{equation}

\bibliographystyle{utphys}
\bibliography{DisContent}

\end{document}